\pdfoutput=1

\documentclass[11pt,twoside,a4paper,cmspaper,final,collab]{cms-tdr}
\def\svnVersion{94404:94406}\def\cmsCernNo{2011-187}\def\cmsCernDate{\today}\def\cmsMessage{Submitted to the Journal of High Energy Physics}

\begin{document}\cmsNoteHeader{FWD-10-005}

\hyphenation{had-ron-i-za-tion}
\hyphenation{cal-or-i-me-ter}
\hyphenation{de-vices}
\RCS$Revision: 48948 $
\RCS$HeadURL: svn+ssh://svn.cern.ch/reps/tdr2/notes/FWD-10-005/trunk/FWD-10-005.tex $
\RCS$Id: FWD-10-005.tex 48948 2011-04-07 08:49:41Z jjhollar $

\cmsNoteHeader{FWD-10-005} 
\title{Exclusive $\Pgg\Pgg\rightarrow\Pgmp\Pgmm$ production in proton-proton collisions at $\sqrt{s} = 7$~TeV}

\date{\today}

\abstract{
A measurement of the exclusive two-photon production of muon pairs in proton-proton collisions at $\sqrt{s}= 7$~TeV,
$\Pp\Pp \rightarrow \Pp \Pgmp\Pgmm \Pp$, is reported using data corresponding to
an integrated luminosity of $40$~pb$^{-1}$. For muon pairs with invariant mass greater than $11.5$~GeV, transverse momentum $p_T(\mu) > 4$~GeV
and pseudorapidity $|\eta(\mu)| < 2.1$, a fit to the dimuon $\pt(\Pgmp\Pgmm)$ distribution results in a measured cross section of
$\sigma(\Pp \rightarrow \Pp \Pgmp\Pgmm \Pp) = 3.38^{+0.58}_{-0.55}~(\mathrm{stat.}) \pm 0.16~(\mathrm{syst.}) \pm 0.14~(\mathrm{lumi.})$~pb, consistent with the theoretical prediction evaluated with the event generator \textsc{Lpair}.
The ratio to the predicted cross section is $0.83^{+0.14}_{-0.13}~(\mathrm{stat.}) \pm 0.04~(\mathrm{syst.}) \pm 0.03~(\mathrm{lumi.})$.
The characteristic distributions of the muon pairs produced via $\Pgg\Pgg$ fusion, such as the muon acoplanarity, the muon
pair invariant mass and transverse momentum agree with those from the theory.
}

\hypersetup{
pdfauthor={CMS Collaboration},
pdftitle={Exclusive photon-photon production of muon pairs in proton-proton collisions at sqrt(s) = 7 TeV},
pdfsubject={CMS},
pdfkeywords={CMS, physics}}

\maketitle

\section{Introduction}

\begin{figure}[h!]
\centering
\includegraphics[width=0.3\textwidth]{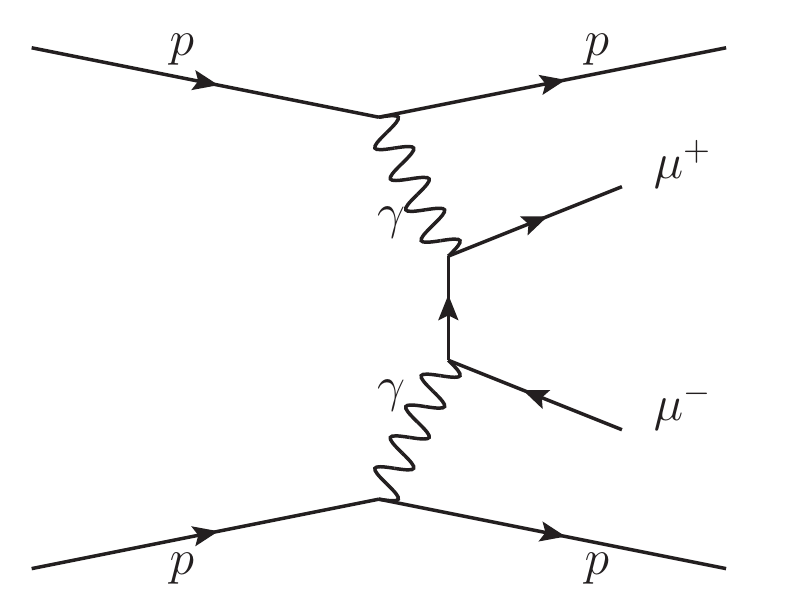}
\includegraphics[width=0.3\textwidth]{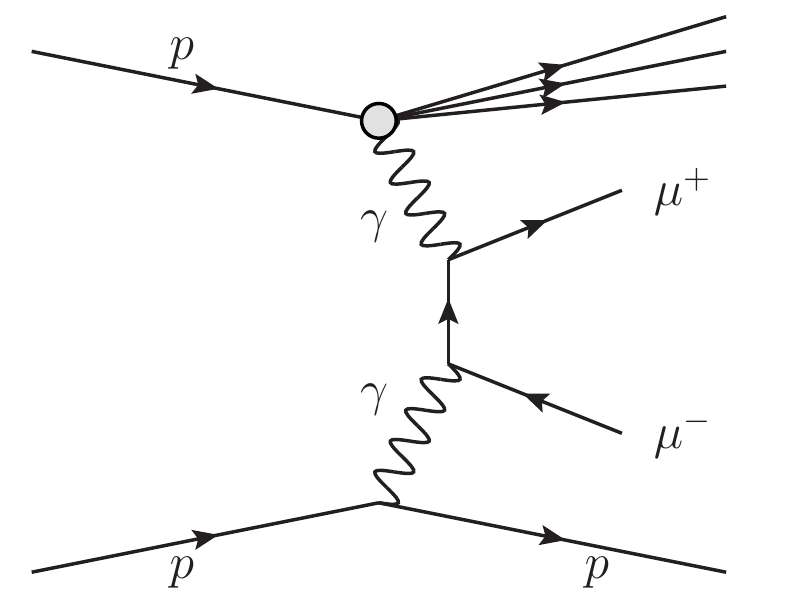}
\includegraphics[width=0.3\textwidth]{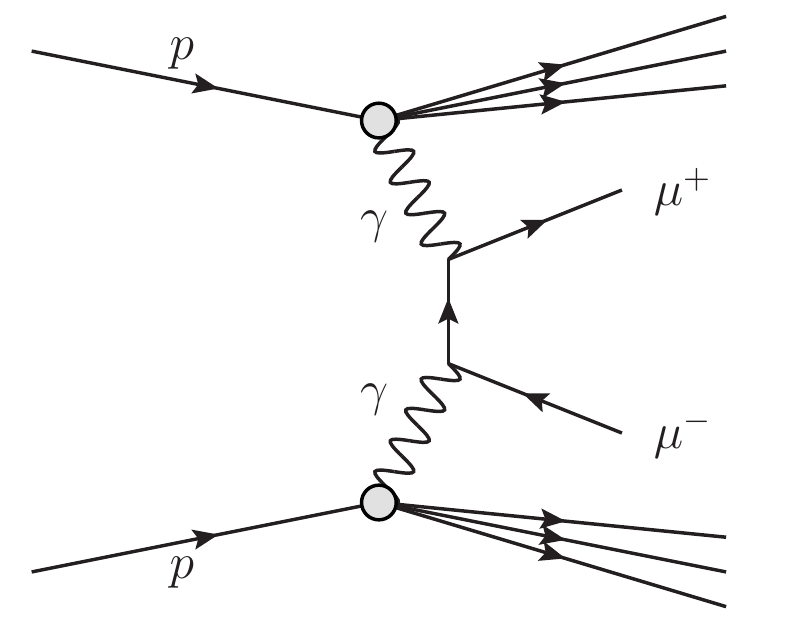}
\caption{ \small{Schematic diagrams for the exclusive and semi-exclusive two-photon production of muon pairs in pp collisions for the elastic (left), single dissociative (center), and double dissociative (right) cases. The three lines in the final state of the center and right plots indicate dissociation of the proton into a low-mass system $N$. }}
\label{fig:feynman}
\end{figure}

The exclusive two-photon production of lepton pairs may be reliably calculated within the framework of quantum
electrodynamics (QED)~\cite{Budnev1} (Fig.~\ref{fig:feynman}), within uncertainties of less than $1\%$ associated with the
proton form factor~\cite{lumi2}. Indeed, detailed theoretical studies have shown that corrections due to hadronic interactions
between the elastically scattered protons are well below 1$\%$ and can be safely neglected \cite{lumi1}. The unique features of this
process, like the extremely small pair transverse momentum and acoplanarity (defined as $1 - |\Delta \phi(\Pgmp\Pgmm)/\pi|$), stem
from the very small virtualities of the exchanged photons.

At the Tevatron, the exclusive two-photon production of electron~\cite{Abulencia,CDFZ} and muon~\cite{CDFchic,CDFZ}
pairs in $\Pp\Pap$ collisions has been measured with the CDF detector. Observations have been made of QED signals,
leading to measurements of exclusive charmonium photoproduction~\cite{CDFchic} and searches for anomalous high-mass
exclusive dilepton production~\cite{CDFZ}. However, all such measurements have very limited numbers of selected events because the data samples
were restricted to single interaction bunch crossings. The higher energies and increased luminosity available at the
Large Hadron Collider (LHC) will allow significant improvements in these measurements, if this limitation can be avoided.
As a result of the small theoretical uncertainties and characteristic kinematic distributions in $\Pgg\Pgg \rightarrow \Pgmp\Pgmm$, this
process has been proposed as a candidate for a complementary absolute calibration of the luminosity of pp collisions \cite{Budnev1,lumi1,lumi2}.

Unless both outgoing protons are detected, the semi-exclusive two-photon production,
involving single or double proton dissociation (Fig.~1, middle and right panels), becomes an
irreducible background that has to be subtracted. The proton-dissociation process is less well determined
theoretically, and in particular requires significant corrections due to proton rescattering.
This effect occurs when there are strong-interaction exchanges between the protons, in addition
to the two-photon interaction. These extra contributions may alter the kinematic distributions of the
final-state muons, and may also produce additional low-momentum hadrons. As a result, the proton-dissociation process has
significantly different kinematic distributions compared to the pure exclusive case, allowing an effective separation
of the signal from this background.

In this paper, we report a measurement of dimuon exclusive production in pp collisions at
$\sqrt{s} = 7$\TeV for the invariant mass of the pair above 11.5\GeV, with each muon having transverse momentum $\pt(\mu) > 4$\GeV
and pseudorapidity $|\eta(\mu)| < 2.1$ (where $\eta$ is defined as $-\ln(\tan(\theta/2))$).
This measurement is based on data collected by the Compact Muon Solenoid (CMS) experiment during the 2010 LHC run, including beam
collisions with multiple interactions in the same bunch crossing (event pileup), and corresponding to an integrated luminosity of 40\pbinv with a relative uncertainty of $4\%$~\cite{LUMIDP}.

The paper is organized as follows. In Section 2, a brief description of the CMS detector is provided. Section 3 describes the
data and samples of simulated events used in the analysis. Section 4 documents the criteria used to select events, and Section 5 the
method used to extract the signal yield from the data. The systematic uncertainties and cross-checks performed are discussed
in Section 6, while Section 7 contains plots comparing the selected events in data and simulation. Finally, the results
of the measurement are given in Section 8 and summarized in Section 9.

\section{The CMS detector}

A detailed description of the CMS experiment
can be found elsewhere~\cite{JINST}.
The central feature of the CMS apparatus is a superconducting
solenoid, of 6\unit{m} internal diameter.  Within the field volume are the
silicon pixel and strip tracker, the crystal electromagnetic
calorimeter, and the brass/scintillator hadronic calorimeter.
Muons are measured in gaseous detectors embedded in the iron
return yoke.  Besides the barrel and endcap detectors, CMS has
extensive forward calorimetry.
CMS uses a right-handed coordinate system, with the origin at the
nominal collision point, the $x$ axis pointing to the center of the
LHC ring, the $y$ axis pointing up (perpendicular to the plane of the LHC ring), and
the $z$ axis along the anticlockwise-beam direction.
The azimuthal angle $\phi$ is measured in the $x$-$y$ plane.
Muons are measured in the window $|\eta|< 2.4$,
with detection planes made using three systems: drift tubes, cathode
strip chambers, and resistive plate chambers.
Thanks to the strong magnetic field, 3.8\unit{T}, and to the
high granularity of the silicon tracker (three layers consisting of
66~million $100\times150\micron^2$ pixels followed by ten microstrip
layers, with strips of pitch between 80 and 180\micron), the $\pt$ of
the muons matched to silicon tracks is measured with a resolution better than $\sim$\,1.5\%, for
$\pt$ less than 100\GeV.
The first level of the CMS trigger system, composed of custom
hardware processors, uses information from the calorimeters and muon
detectors to select (in less than 1\mus) the most interesting
events. The High Level Trigger processor farm further decreases the event rate
from 50--100\unit{kHz} to a few hundred Hz, before data storage.

\section{Simulated Samples}

The \textsc{Lpair 4.0} event generator~\cite{Vermaseren,Baranov} is used to produce simulated samples
of two-photon production of muon pairs. The generator uses full leading-order QED matrix elements, and
the cross sections for the exclusive events depend on the proton electromagnetic form-factors
to account for the distribution of charge within the proton. For proton dissociation, the cross sections depend on
the proton structure function. In order to simulate the fragmentation of the dissociated proton into a low-mass
system $N$, the \textsc{Lund} model shower routine~\cite{lund} implemented in the \textsc{JetSet} software~\cite{jetset} is
used with two different structure functions. For masses of the dissociating system $m_N < 2$~GeV and photon
virtualities $Q^2 < 5\GeV^2$, the Brasse ``cluster"
fragmentation is chosen~\cite{brasse}, while for the other cases the Suri-Yenni ``string"
fragmentation is applied~\cite{suri}. In the first case, the low-mass system $N$ mostly
decays to a $\PgD^+$ or $\PgD^{++}$ resonance, which results in low-multiplicity states.
In the second case, the high-mass system usually decays to a variety of
resonances ($\PgD$, $\rho$, $\PgO$, $\eta$, $\PK$), which produce a large number of forward protons,
pions, neutrons and photons. No corrections are applied to account for rescattering effects.
In general, these effects are expected to increase with the transverse momentum of the muon pair, modifying
the slope of the $\pt^{2}(\Pgmp\Pgmm)$ distribution~\cite{lumi1}.

The inclusive Drell--Yan (DY) and quantum chromodynamic (QCD) dimuon backgrounds are simulated
with \PYTHIA v. 6.422~\cite{pythia6}, using the Z2 underlying event tune~\cite{tuneZ2}. All these samples are then passed through
the full \GEANTfour detector simulation~\cite{geant} in order to determine the signal and background
efficiencies after all selection criteria are applied.

\section{Event selection}

The analysis uses a sample of pp collisions at $\sqrt{s}=7$~TeV, collected during 2010 at the LHC and
corresponding to an integrated luminosity of 40 pb$^{-1}$. The sample includes 36 pb$^{-1}$ of data passing the standard
CMS quality criteria for all detector subsystems, and 4 pb$^{-1}$ in which the quality criteria are satisfied for the tracking
and muon systems used in the analysis. From the sample of triggered events, the presence of two
reconstructed muons is required. Then the exclusivity selection is performed to keep only events with a vertex having no tracks other
than those from the two muons. Finally, the signal muons are required to satisfy identification criteria, and kinematic constraints
are imposed using their four-momentum. All selection steps are described in the following sections.

\subsection{Trigger and muon reconstruction}

Events are selected online by triggers requiring the presence of two muons with a minimum $\pt$ of 3\GeV. No requirement on
the charge of the muons is applied at the trigger level. Muons are reconstructed offline by combining information from the muon chambers
with that on charged-particle tracks reconstructed in the silicon tracker~\cite{CRAFTMUONS}, and events with a pair of oppositely
charged muons are selected.

\subsection{Vertex and track exclusivity selection}

With single interactions, the exclusive signal is characterized by the presence of two muons, no additional tracks, and no activity
above the noise threshold in the calorimeters. The presence of additional interactions in the same bunch crossing
will spoil this signature by producing additional tracks and energy deposits in the calorimeters. In the 2010
data, less than 20\% of the total luminosity was estimated to have been collected from bunch crossings where only a single
interaction look place, leading to a significant decrease in signal efficiency if the conditions of no extra tracks or
calorimeter energy are required.

The selection of exclusive events is therefore applied using the pixel and silicon tracker only, since the primary vertex
reconstruction \cite{bib-trackingefficiency,bib-vertex} allows discrimination
between different interactions within the same bunch crossing. The selection requires a valid vertex, reconstructed using an
adaptive vertex fit to charged-particle tracks clustered in $z$~\cite{bib-adaptivevtx,bib-vertex}, with exactly two
muons and no other associated tracks, and vertex fit probability greater than 0.1\%. The dimuon vertex is further required to have coordinates
consistent with a collision in CMS, with a longitudinal displacement of less than 24\unit{cm} and a transverse displacement of
less than 0.1\unit{cm}.

In order to reduce the background from inclusive DY and QCD dimuon production, which typically have many tracks originating
from the same vertex as a prompt muon pair, the dimuon vertex is required to be separated in three dimensions by more than 2\unit{mm} from
any additional tracks in the event. This value is selected to optimize the signal efficiency and background rejection found in
events triggered only by the presence of colliding bunches (``zero-bias" events), and in DY Monte Carlo simulation. For the zero-bias data, this
is accomplished by introducing an artificial additional dimuon vertex into each event as a proxy for an exclusive dimuon interaction.
Thus, in this study, beam crossings with no real vertex present are counted as ``single vertex" events, and crossings with one real vertex
are counted as having an additional pileup event.

\begin{figure}[h!]
\centering
\includegraphics[width=0.5\textwidth]{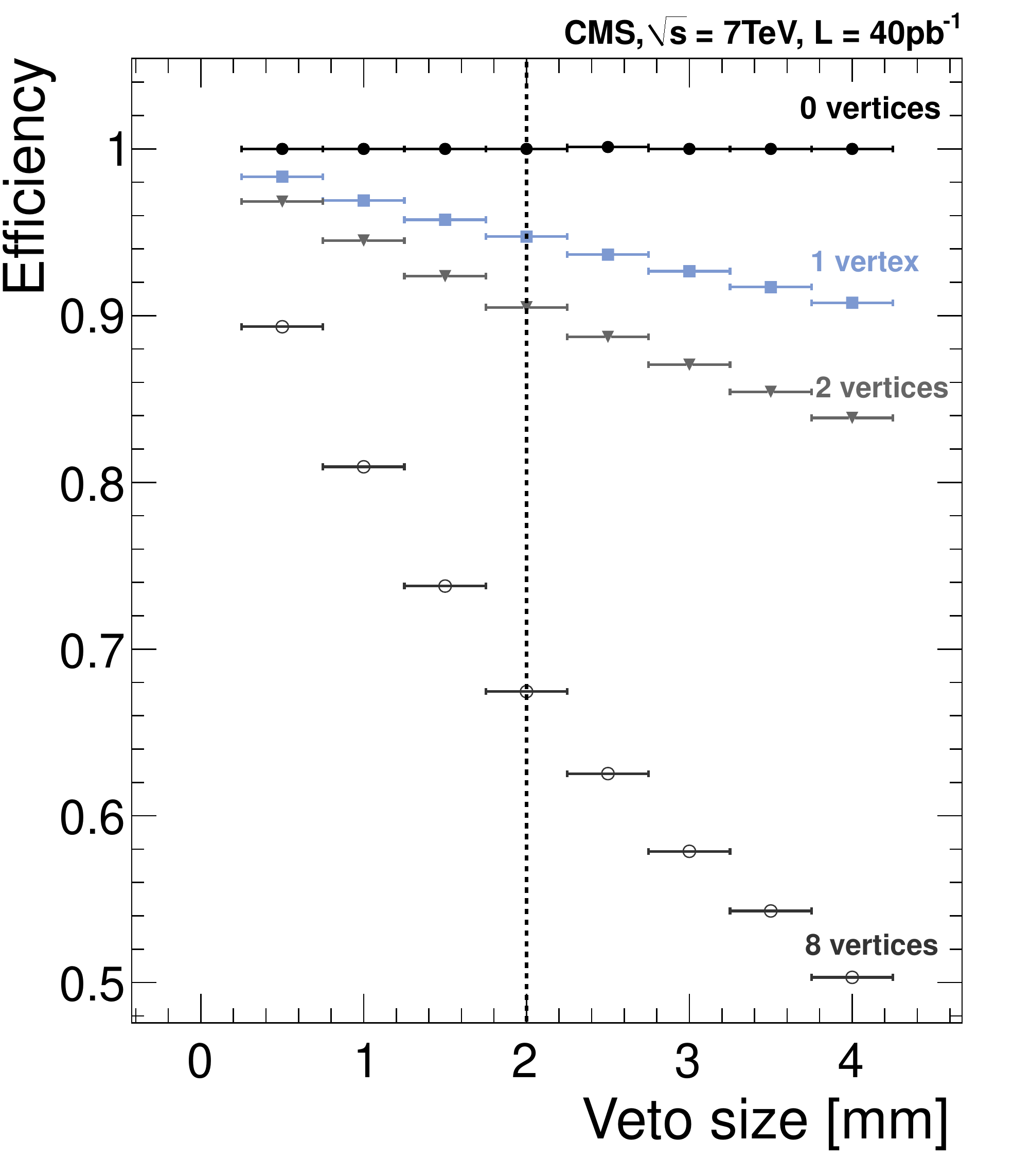}
\caption{ \small{Efficiency of the zero extra tracks selection vs. distance to closest
track computed with the artificial vertex method in zero-bias data. The points correspond to
events with 0, 1, 2, and 8 real vertices in the event. Events to the right of the vertical dashed line are selected.
The vertical error bars are negligible.} }
\label{fig:vertexefficiency}
\end{figure}

\begin{figure}[h!]
\centering
\includegraphics[width=0.5\textwidth]{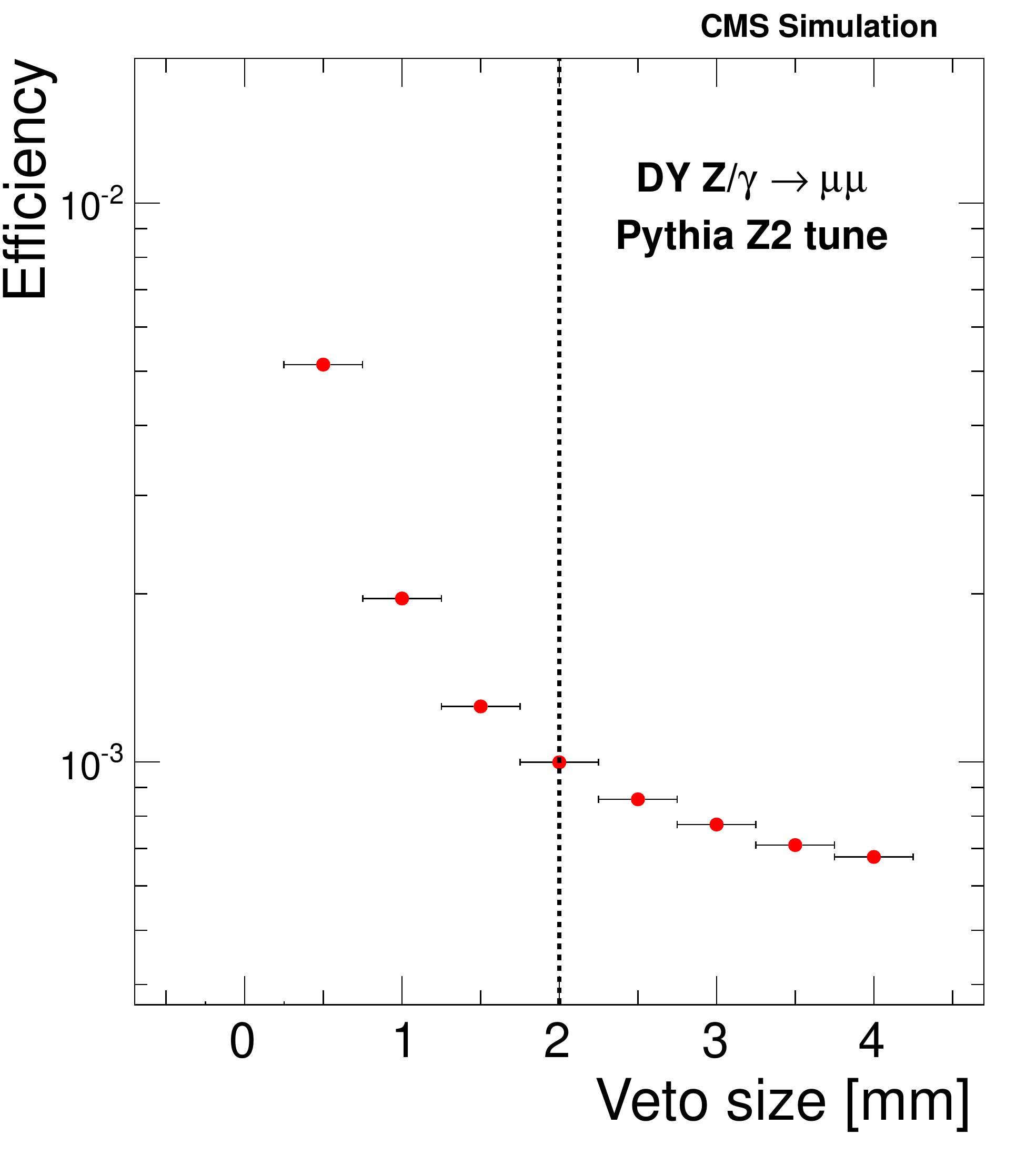}
\caption{ \small{Efficiency of the zero extra tracks selection vs. distance to closest track computed for DY events in simulation.
Events to the right of the vertical dashed line are selected. The vertical error bars are negligible.}}
\label{fig:dyefficiency}
\end{figure}

The effects of the track veto on the signal efficiency and on the efficiency for misidentifying background as signal are studied as a function
of the distance to the closest track for the zero-bias sample and DY background (Figs.~\ref{fig:vertexefficiency} and \ref{fig:dyefficiency}).
With no extra vertices in the zero-bias events, the efficiency approaches $100\%$ as expected for events with no pileup. With the addition
of overlap events, the efficiency decreases, falling to $\sim60\%$ with 8 extra vertices reconstructed. In the full data sample the
average number of extra vertices is 2.1, with less than $10\%$ of events having 4 or more extra vertices. The efficiency for
misidentifying the DY background as signal increases sharply for distances less than 1\unit{mm}, consistent with the resolution of the
single-track impact parameter in the longitudinal direction~\cite{bib-vertex}.

\subsection{Muon identification}

Each muon of the pair is required to pass a ``tight" muon selection~\cite{MUPOGICHEPPAS}. This selection consists of
requesting that the reconstructed muon have at least one hit in the pixel detector, at least 10 hits in the silicon strip tracker, and segments
reconstructed in at least two muon detection planes. In addition, a global fit to the combined information from the tracker and
muon systems must include at least one muon chamber hit, and have a $\chi^{2}$ per degree of freedom of less than 10.

\subsection{Kinematic selection}

In order to minimize the systematic uncertainties related to the knowledge of the low-$p_{\mathrm{T}}$ and large-$\eta$ muon efficiencies, only muons
with $p_{\mathrm{T}} > 4$~GeV and $|\eta| < 2.1$ are selected. The $p_{\mathrm{T}}$ and $|\eta|$ requirements retain muon pairs from
exclusive photoproduction of upsilon mesons, $\gamma \mathrm{p} \rightarrow \Upsilon \mathrm{p} \rightarrow \mu^{+} \mu^{-} \mathrm{p}$.
This process occurs when a photon emitted from one proton fluctuates into a $q\overline{q}$ pair, which interacts with the second proton via a
color-singlet exchange. This contribution is removed by requiring that the muons have an invariant mass $m(\mu^{+}\mu^{-})>  11.5$~GeV.

In order to suppress further the proton dissociation background, the muon
pair is required to be back-to-back in azimuthal angle ($1 - |\Delta \phi(\Pgmp\Pgmm)/\pi| < 0.1$) and balanced
in the scalar difference in the $\pt$ of the two muons ($|\Delta \pt(\Pgmp\Pgmm)| < 1.0$~GeV). A possible
contamination could arise from cosmic-ray muons, which
would produce a signature similar to the exclusive $\Pgg\Pgg\rightarrow\mu^+\mu^-$ signal. The
three-dimensional opening angle of the pair, defined as the arccosine of the normalized scalar product
of the muon momentum vectors, is therefore required to be smaller than 0.95~$\pi$, to reduce any contribution from cosmic-ray muons.

The effect of each step of the selection on the data and simulated signal and background samples is shown in Table~\ref{tab:eff}.
After all selection criteria are applied, 148 events remain, where from simulation, approximately half are expected to originate
from elastic production. The number of events selected in data is below the expectation from simulation, with an observed
yield that is roughly $80\%$ of the sum of simulated signal and background processes. The deficit could be caused by a
lower-than-expected signal yield, or by a smaller proton-dissociation contribution than expected from simulation.

\begin{table}[h!]
\begin{center}
\caption{Number of events selected in data and number of signal and background events expected from simulation at each selection step for
an integrated luminosity of 40\pbinv. The last column is the number of events expected from the sum of the signal, DY, and proton dissociation
backgrounds in the simulation. The relative statistical uncertainty on the sum of simulated signal and background samples in each row
is $\leq 0.5\%$. The contribution from exclusive resonance production of $\PgU$ or $\chi_{b}$ mesons is not simulated, and thus contributes
only to the data column before requiring $m(\Pgmp\Pgmm) > 11.5$\GeV. For entries in the line ``Muon ID" and below, the simulation
is corrected for effects related to event pileup, muon identification, trigger, and tracking efficiencies, as described in the text.}
\begin{tabular}{l|c|cccc|c}
\hline
Selection & Data & Signal & Single-\Pp diss. & Double-\Pp diss. & DY & Total\\
\hline
Vertex and track-exclusivity & 921 & 247  & 437 & 197 & 56  & 937 \\
Muon ID                        & 724 & 193  & 336 & 160 & 53  & 741 \\
$\pt >4$~GeV, $|\eta| < 2.1$ & 438 & 132  & 241 & 106 & 20  & 499 \\
$m(\Pgmp\Pgmm) > 11.5$~GeV         & 270 &  95  & 187 &  86 & 13  & 380 \\
$3D$ angle $< 0.95 \pi$        & 257 &  87  & 178 &  83 & 12  & 361 \\
$1 -|\Delta \phi/\pi| < 0.1$   & 203 &  87  & 126 &  41 &  8   & 263 \\
$|\Delta \pt| < 1.0$~GeV     & 148 &  86  &  79 &  16 &  3  & 184 \\
\hline
\end{tabular}
\label{tab:eff}
\end{center}
\end{table}

\section{Signal extraction}

\subsection{Efficiency corrections}

A correction is applied to account for the presence of extra proton-proton interactions in the same bunch crossing as a signal event. These
pileup interactions will result in an inefficiency if they produce a track with a position within the nominal $2$\unit{mm} veto distance around
the dimuon vertex. This effect is studied in zero-bias data using the method described in Section~$4.2$.
The nominal 2\unit{mm} veto is then applied around the dimuon vertex, and the event is accepted if no tracks fall within the veto distance.
The efficiency is measured as a function of the instantaneous luminosity per colliding bunch. The average efficiency is calculated
based on the instantaneous luminosities to be $92.29\%$ for the full 2010 data set, with negligible statistical uncertainty.

The trigger, tracking, and offline muon selection efficiencies are each obtained from the tag-and-probe~\cite{MUPOGICHEPPAS,CMSWZ} method by using
samples of inclusive $\PJGy \rightarrow \Pgmp\Pgmm$ and \cPZ$ \rightarrow \Pgmp\Pgmm$ events from data and Monte Carlo simulation.
These control samples are triggered on one muon such that the other muon is unbiased with respect to the efficiency to be measured.
For $\pt < 20$\GeV muons from \PJGy decays are used, while above 20\GeV muons from \cPZ\ decays are used.
The trigger and offline muon selection efficiencies are measured using $\PJGy$ events by requiring a muon tag that, when combined with a
track reconstructed using only the silicon detectors, is consistent with a $\PJGy$. These efficiencies
are measured in bins in $\pt$ and  $\eta$, separately for the two muon charges. The tracking efficiency is measured similarly
using \PJGy events by requiring a muon tag that, when combined with a muon reconstructed using only the muon systems, is
consistent with a \PJGy. The tracking efficiency is then measured on the unbiased muon probe. In contrast to the trigger and
muon identification efficiencies, the tracking efficiency is measured in data and Monte Carlo simulation
averaged over $|\eta| < 2.1$ and $\pt > 4\GeV$, and is taken to be uncorrelated between the two tracks. The resulting
ratio of efficiencies in data and simulation for the pair $(99.18 \pm 0.14)\%$ is applied as a correction to the efficiency.

The effect of the vertexing efficiency is studied both in inclusive dimuon data and signal simulation, by performing an independent
selection of all muon pairs with a longitudinal separation of less than 0.5\unit{mm}. A Kalman filter~\cite{Kalman} algorithm is then
applied to estimate the best position of the dimuon vertex, without using information from any tracks other than the two muons.
Among events with a valid dimuon vertex and for which no additional tracks exist within 2\unit{mm} in $z$, the efficiency for the default
adaptive vertex fitter to reconstruct a primary vertex with only two muons attached and matching with the Kalman vertex is computed. The
ratio of the vertexing efficiency in data to that in simulation is $99.97\%$, and therefore no correction is applied.

\subsection{Maximum likelihood fit}

\begin{figure}[h!]
\centering
\includegraphics[width=0.5\textwidth]{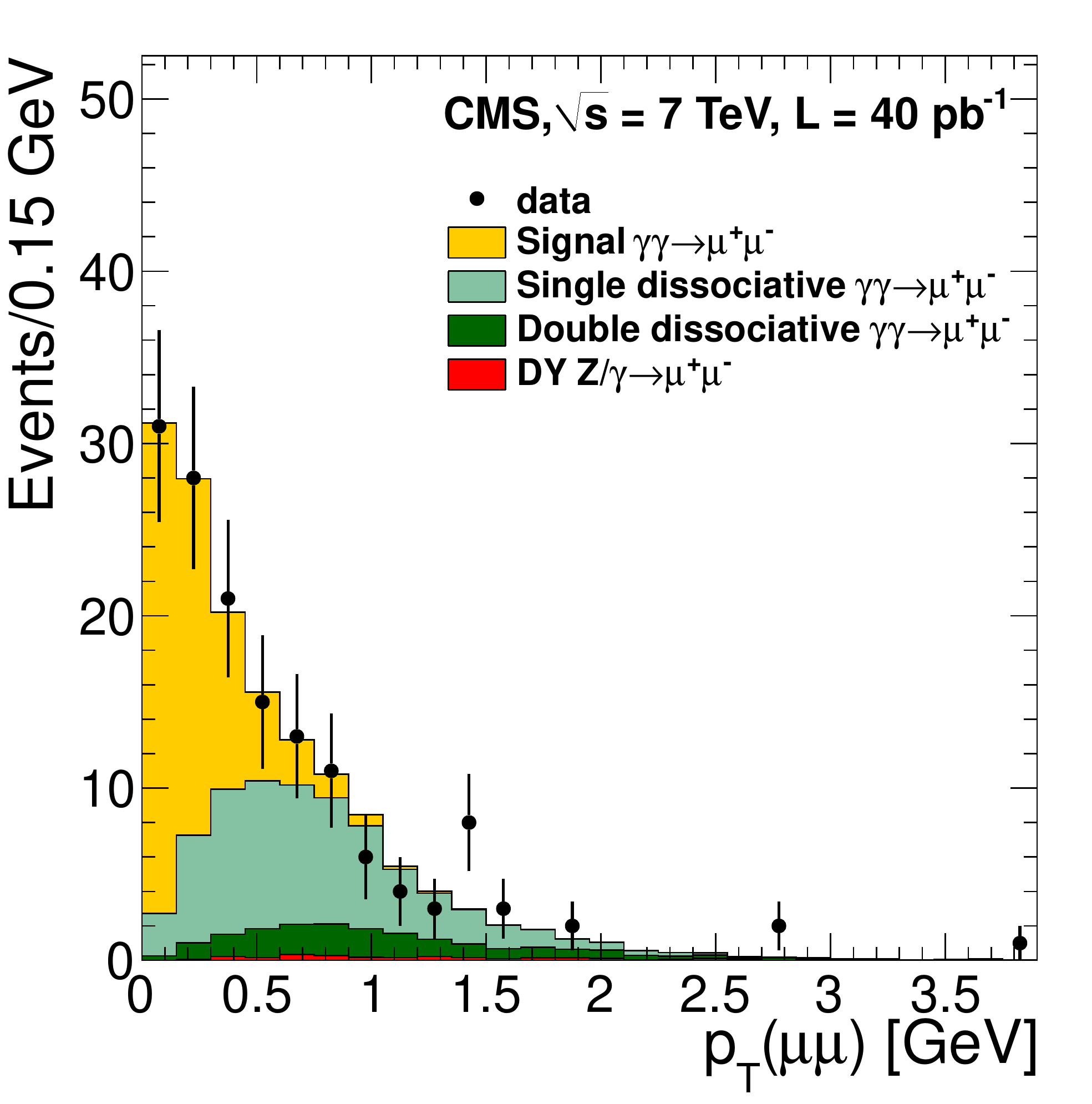}
\caption{Distribution of $\pt(\Pgmp\Pgmm)$ for the selected sample. Data are shown as points with statistical error bars. The histograms
represent the simulated signal (yellow), single (light green) and double (dark green) proton dissociative backgrounds, and DY (red). The yields are
determined from a fit using the distributions from simulation.}
\label{fig:ptfitresult}
\end{figure}

The elastic $\Pp\Pp \rightarrow \Pp\Pgmp\Pgmm\Pp$ contribution is extracted by performing a binned maximum-likelihood fit to the
measured $\pt(\Pgmp\Pgmm)$ distribution. Shapes from Monte Carlo simulation are used for the signal, single-proton dissociation,
double-proton dissociation, and DY contributions, with all corrections described in Section 4.4 applied.

Three parameters are determined from the fit: the elastic signal yield relative to the \textsc{Lpair} prediction for an integrated luminosity of 40\pbinv
($R_{El-El}$), the single-proton dissociation yield relative to the \textsc{Lpair} single-proton dissociation prediction for 40\pbinv ($R_{diss-El}$),
and an exponential modification factor for the shape of the $\pt$ distribution, characterized by the parameter $a$.
The modification parameter is included to account for possible rescattering effects not included in the simulation, as described in Section 3.
Given the small number of events expected in 40\pbinv, the double-proton dissociation and DY contributions cannot be treated as free parameters
and are fixed from simulation to their predicted values. The contribution from exclusive $\Pgg \Pgg \rightarrow \tau^{+}\tau^{-}$ production is
estimated to be 0.1 events from the simulation, and is neglected.

\begin{figure}[h!]
\centering
\includegraphics[width=\textwidth]{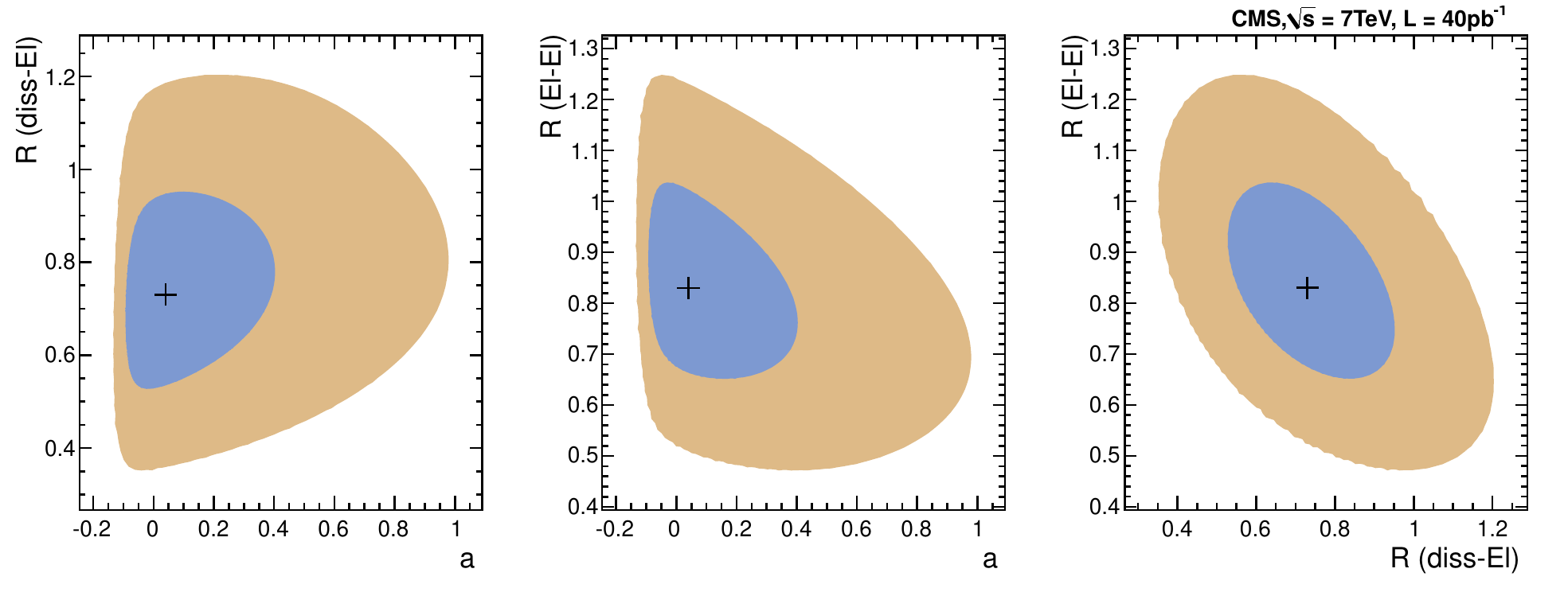}
\caption{One and two standard-deviation contours in the plane of fitted parameters for the proton-dissociation
yield ratio vs. modification parameter $a$ (left), the data-theory signal ratio vs. modification parameter $a$ (center),
and the data-theory signal ratio vs. proton-dissociation yield ratio (right). The contours represent 39.3\% and 86.5\% confidence regions, where
the cross indicates the best-fit point.}
\label{fig:outputfitcontours}
\end{figure}

The $\pt(\Pgmp\Pgmm)$ distribution in data is shown overlaid with the result of the fit to the shapes from Monte Carlo simulation
in Fig.~\ref{fig:ptfitresult}. The result from the best fit to the data is:

\begin{eqnarray}
\textrm{data-theory signal ratio:}                      & R_{El-El} =0.83^{+0.14}_{-0.13} \nonumber; \\
\textrm{single-proton dissociation yield ratio:}        & R_{diss-El} =0.73^{+0.16}_{-0.14} \nonumber; \\
\textrm{modification parameter:}                        & a=0.04^{+0.23}_{-0.14} {\GeV}^{-2} \nonumber, \\
\end{eqnarray}

with asymmetric statistical uncertainties computed using \textsc{minos}~\cite{minuit}. The corresponding value of the signal cross section is
$3.38^{+0.58}_{-0.55}~(\text{stat.})$\unit{pb}. The resulting 1$\sigma$ and 2$\sigma$ contours projected onto each pair of fit
variables are displayed in Fig.~\ref{fig:outputfitcontours}. For any values of the proton dissociation ratio and slope within the
1$\sigma$ contour, the extreme values of the data-theory signal ratio are 0.64 and 1.03. The upper value of 1.03 for the signal
ratio would correspond to the single-proton dissociation component having a ratio to the prediction of approximately 0.65.
The best fit does not require a significant modification parameter. However, the statistical uncertainty on this parameter is chosen to play
the role of a non-negligible systematic uncertainty, to take account of the neglect of the rescattering effects in the simulation.

As a cross-check, a fit to the $1 - |\Delta \phi(\Pgmp\Pgmm)/\pi|$ distribution is performed, with the signal and single-proton dissociation yields as free parameters, and the shape of the single-proton dissociation component
fixed from the simulation. The resulting value of the data-theory signal ratio is $0.81^{+0.14}_{-0.13}$, consistent with the nominal fit result.

The central values of the signal and single-proton dissociation yields from the fit are both below the mean number expected for 40\pbinv,
consistent with the deficit shown in Table 1. This is investigated by repeating the fit, first with the $\Delta \pt(\Pgmp\Pgmm) < 1.0$\GeV requirement
removed, and then with both the $\Delta \pt(\Pgmp\Pgmm) < 1.0$\GeV and $1 -|\Delta \phi(\Pgmp\Pgmm)/\pi| < 0.1$ selections removed.
From simulation this is expected to have negligible effect on the signal efficiency, while enhancing the background. The double-proton dissociation
and DY contributions in particular are expected to be small with the nominal selection, but their sum becomes comparable in size to the
signal with the $\Delta \pt(\Pgmp\Pgmm)$ and $1 -|\Delta \phi(\Pgmp\Pgmm)/\pi|$ requirements removed.

\begin{figure}[h!]
\centering
\includegraphics[width=0.4\textwidth]{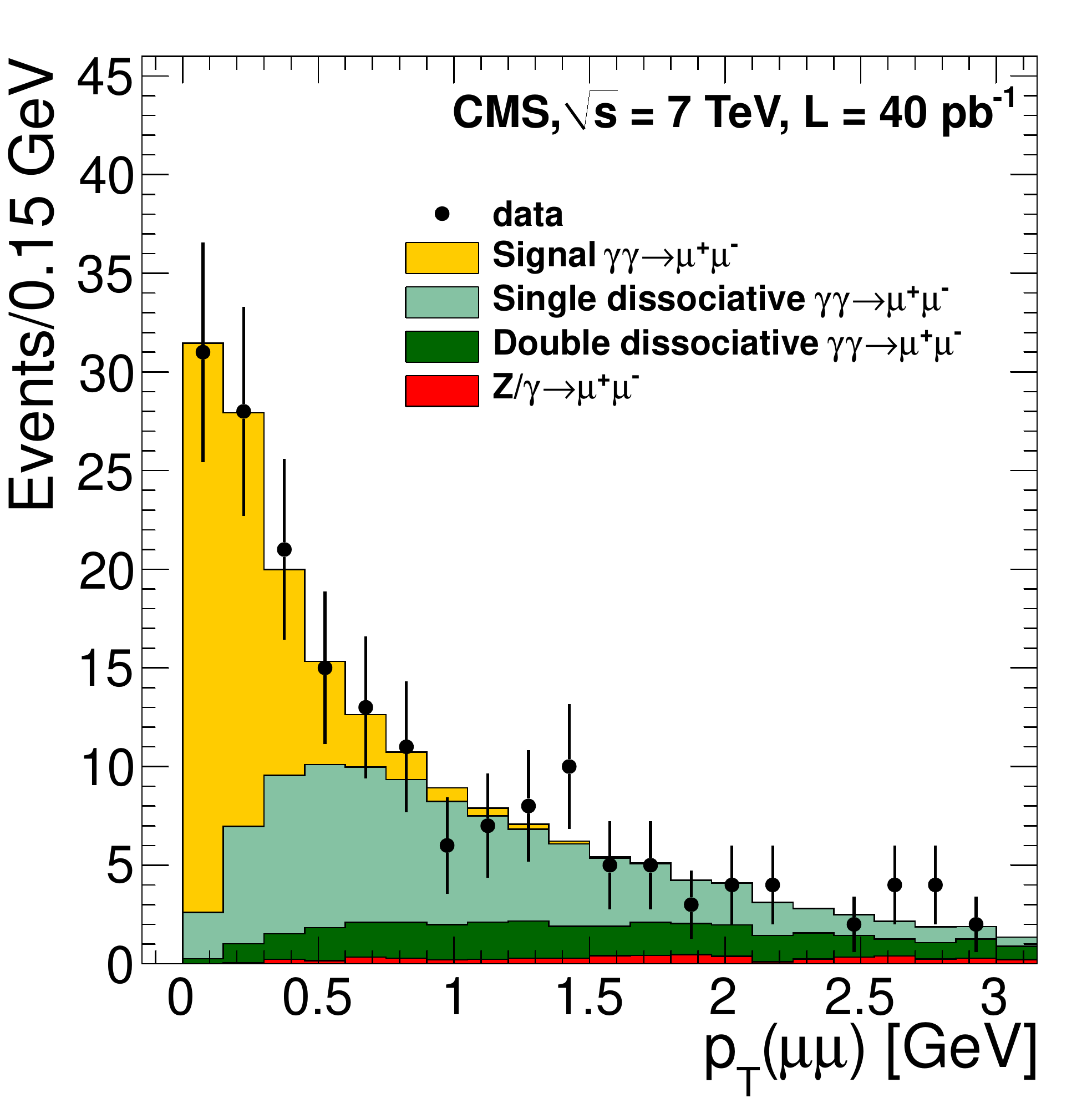}
\includegraphics[width=0.4\textwidth]{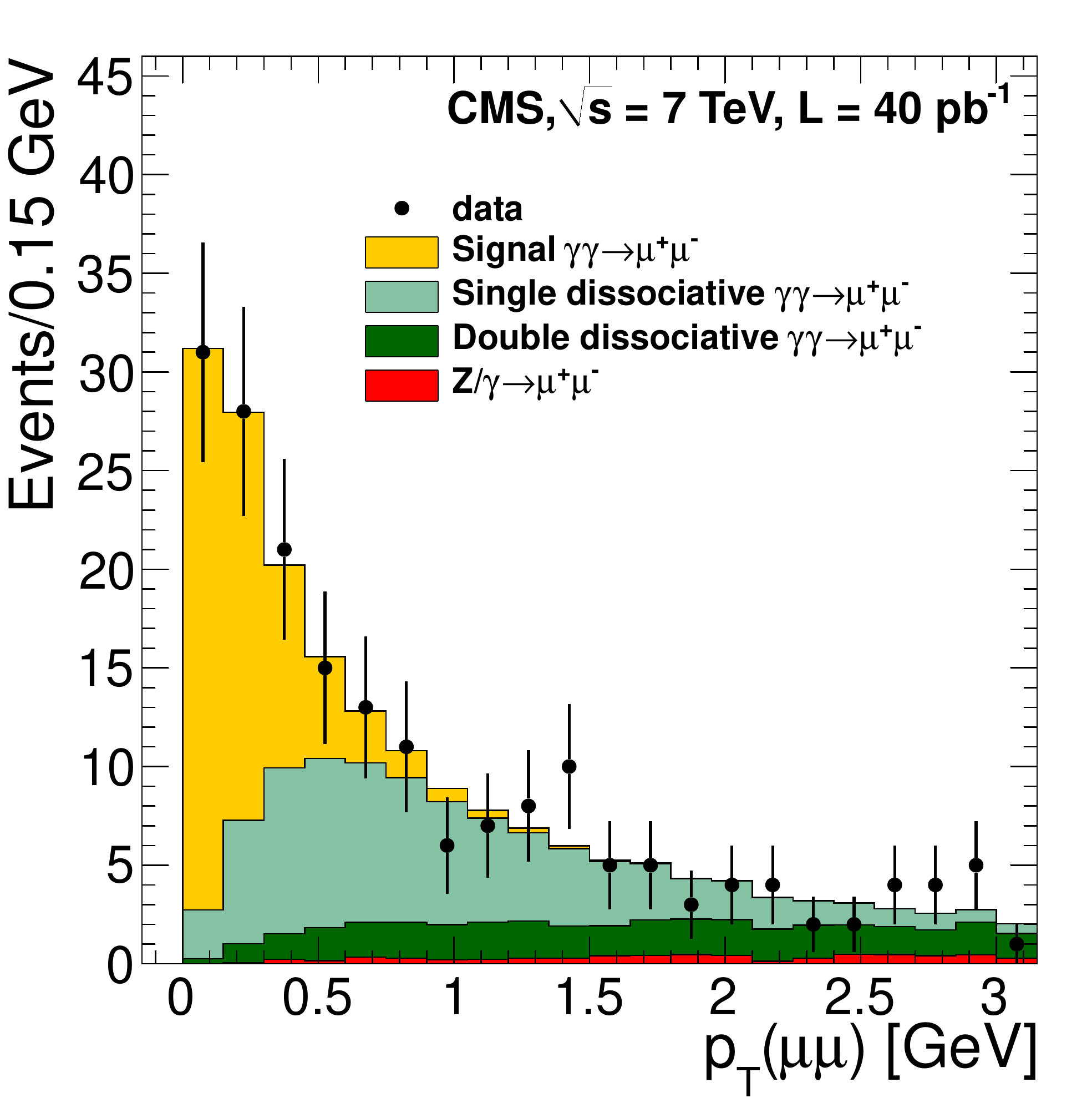}
\caption{Result of fit to the $\pt(\Pgmp\Pgmm)$ distribution with requirements on $|\Delta \pt(\Pgmp\Pgmm)|$ (left) and on both $|\Delta \pt(\Pgmp\Pgmm)|$ and
$1 -|\Delta \phi(\Pgmp\Pgmm)/\pi|$ (right) removed. The points with error bars represent the data. The histograms are the result of fitting
the simulated distributions to the data.}
\label{fig:ptfitplotsloosecuts}
\end{figure}

The fits to the data with these looser selection requirements are shown in Fig.~\ref{fig:ptfitplotsloosecuts}, and the resulting best-fit
yields for the signal and single-proton dissociation are shown in Table~\ref{tab:fitloosecuts}; the single-proton dissociation yield is observed
to be significantly lower relative to the prediction with the looser selections. In all variations, the normalizations of the double-proton
dissociation and DY yields are fixed, although the double-dissociation contribution is expected to be significant at large
$\pt(\Pgmp\Pgmm)$ with the looser selection. With additional data, a more precise comparison of the single and double
dissociation yields to the theoretical expectation may be made. In spite of the lower single-proton dissociation yield, the
data-theory ratio for the signal is stable in all three variations.

\begin{table}[h!]
\begin{center}
\caption{Best-fit values of $R_{El-El}$ and $R_{diss-El}$ for the nominal selection, and with the requirements on $|\Delta \pt(\Pgmp\Pgmm)|$ and $1 -|\Delta \phi(\Pgmp\Pgmm)/\pi|$ removed.}
\renewcommand{\arraystretch}{1.2} 
\begin{tabular}{l|c|c}
\hline
Selection                                     & $R_{El-El}$ & $R_{diss-El}$ \\
\hline
All selection criteria applied                & $0.83^{+0.14}_{-0.13}$ & $0.73^{+0.16}_{-0.14}$ \\\hline
No $|\Delta \pt|$ requirement                           & $0.82^{+0.13}_{-0.13}$ & $0.63^{+0.11}_{-0.10}$ \\\hline
Both $|\Delta \pt|$ and $1 -|\Delta \phi/\pi|$ requirements removed & $0.81^{+0.13}_{-0.13}$ & $0.45^{+0.08}_{-0.07}$ \\\hline

\hline
\end{tabular}
\label{tab:fitloosecuts}
\end{center}
\end{table}

\section{Control plots}

\begin{figure}[h!]
\centering
\includegraphics[angle=0,width=6.5cm]{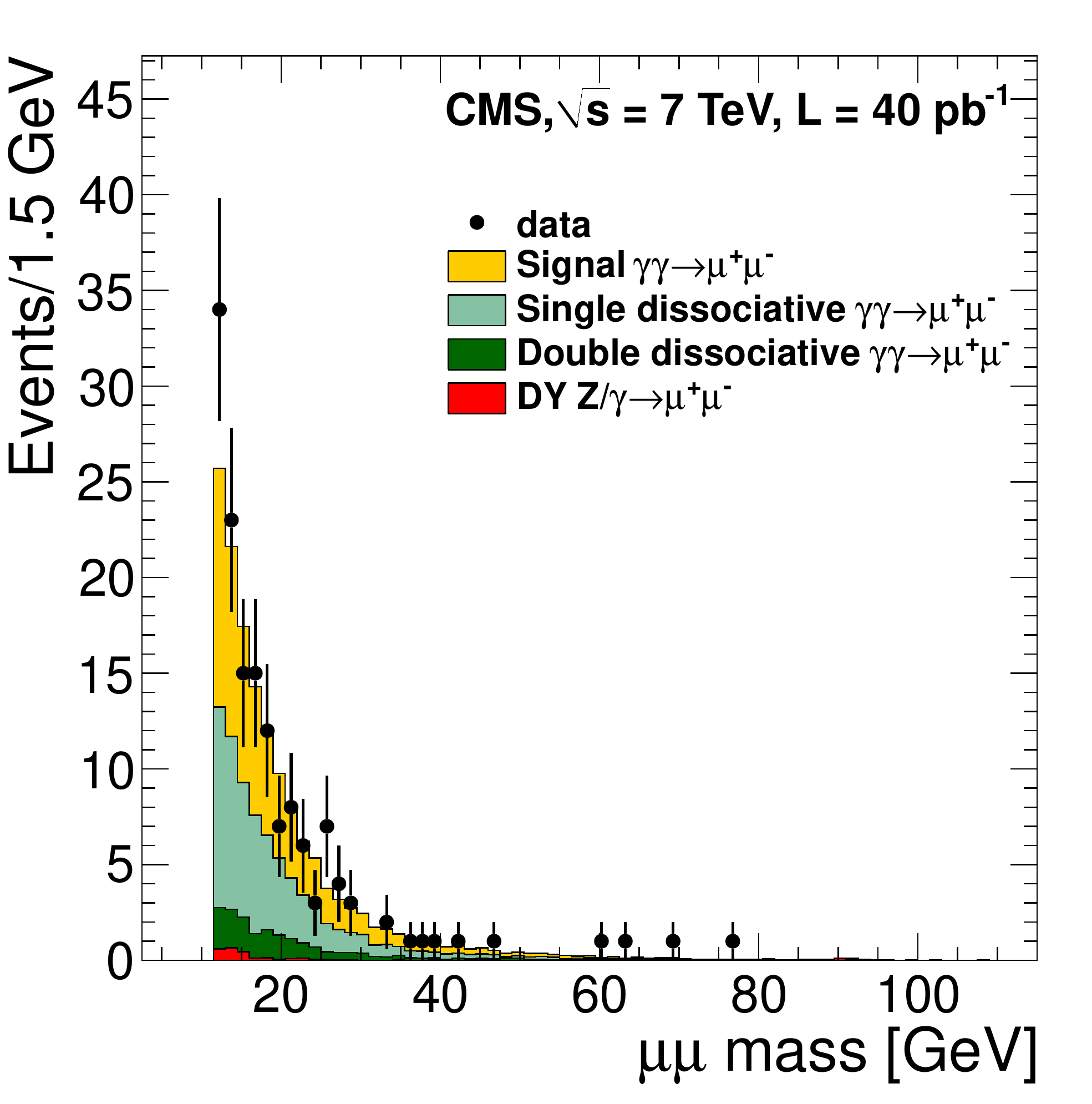}
\includegraphics[angle=0,width=6.5cm]{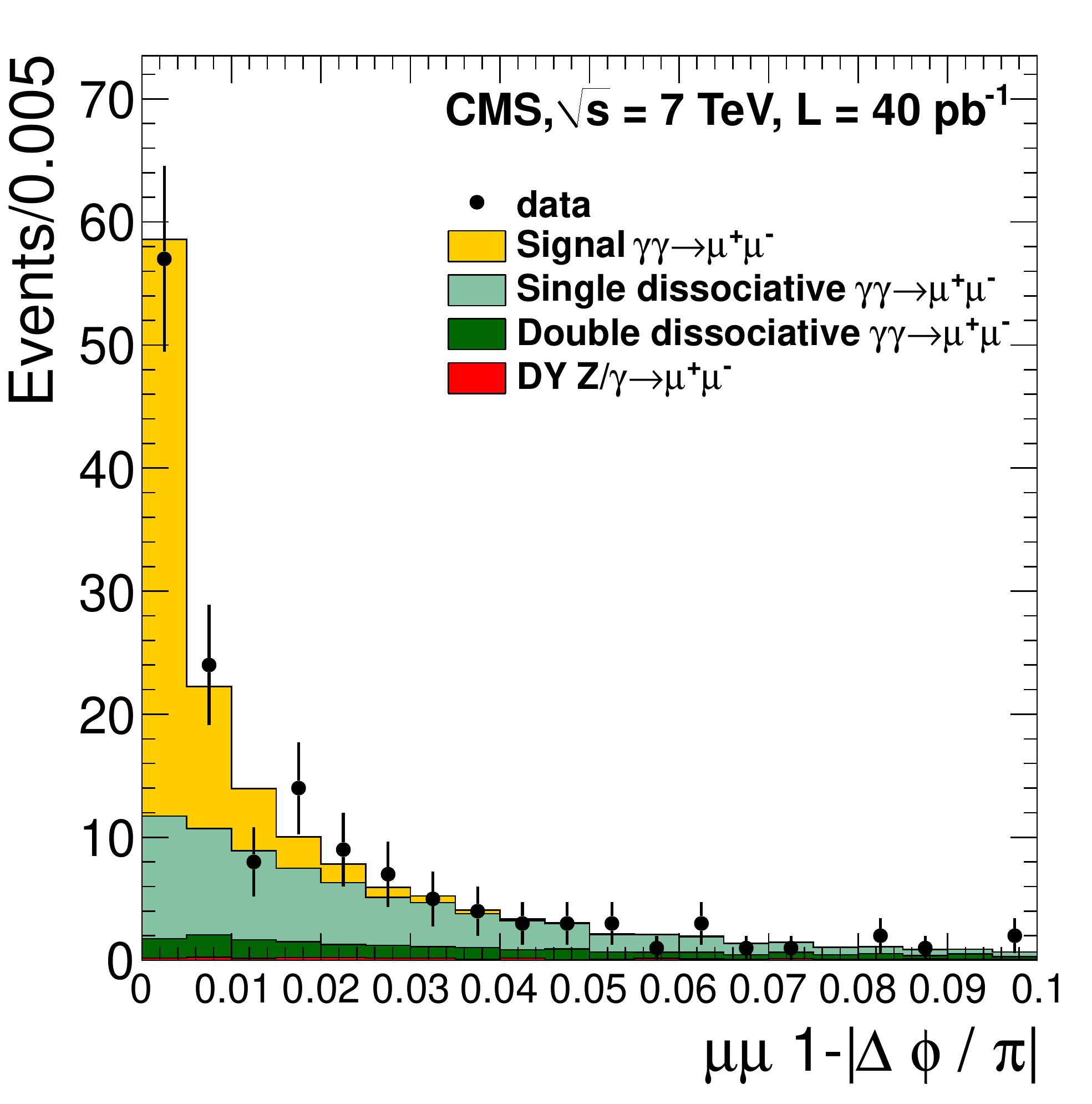}
\caption{Muon pair invariant mass spectrum (left) and acoplanarity (right), with all selection criteria applied and the
simulation normalized to the best-fit value. Data are shown as points with statistical error bars, while the histograms
represent the simulated signal (yellow), single (light green) and double (dark green) proton dissociative backgrounds, and DY (red). }
\label{fig:finalmassplot}
\end{figure}

The dimuon invariant mass and acoplanarity distributions for events passing all selection criteria listed in Table~\ref{tab:eff}
are shown in Fig.~\ref{fig:finalmassplot}, with the simulation predictions normalized to the best-fit signal and background yields. The
event with the largest invariant mass has $m(\Pgmp\Pgmm) = 76\GeV$. No events consistent with
$\cPZ \rightarrow \Pgmp\Pgmm$ are observed. This is expected for exclusive production, since the
$\Pgg \Pgg \rightarrow \cPZ$ process is forbidden at tree-level.

\begin{figure}[h!]
\centering
\includegraphics[width=0.4\textwidth]{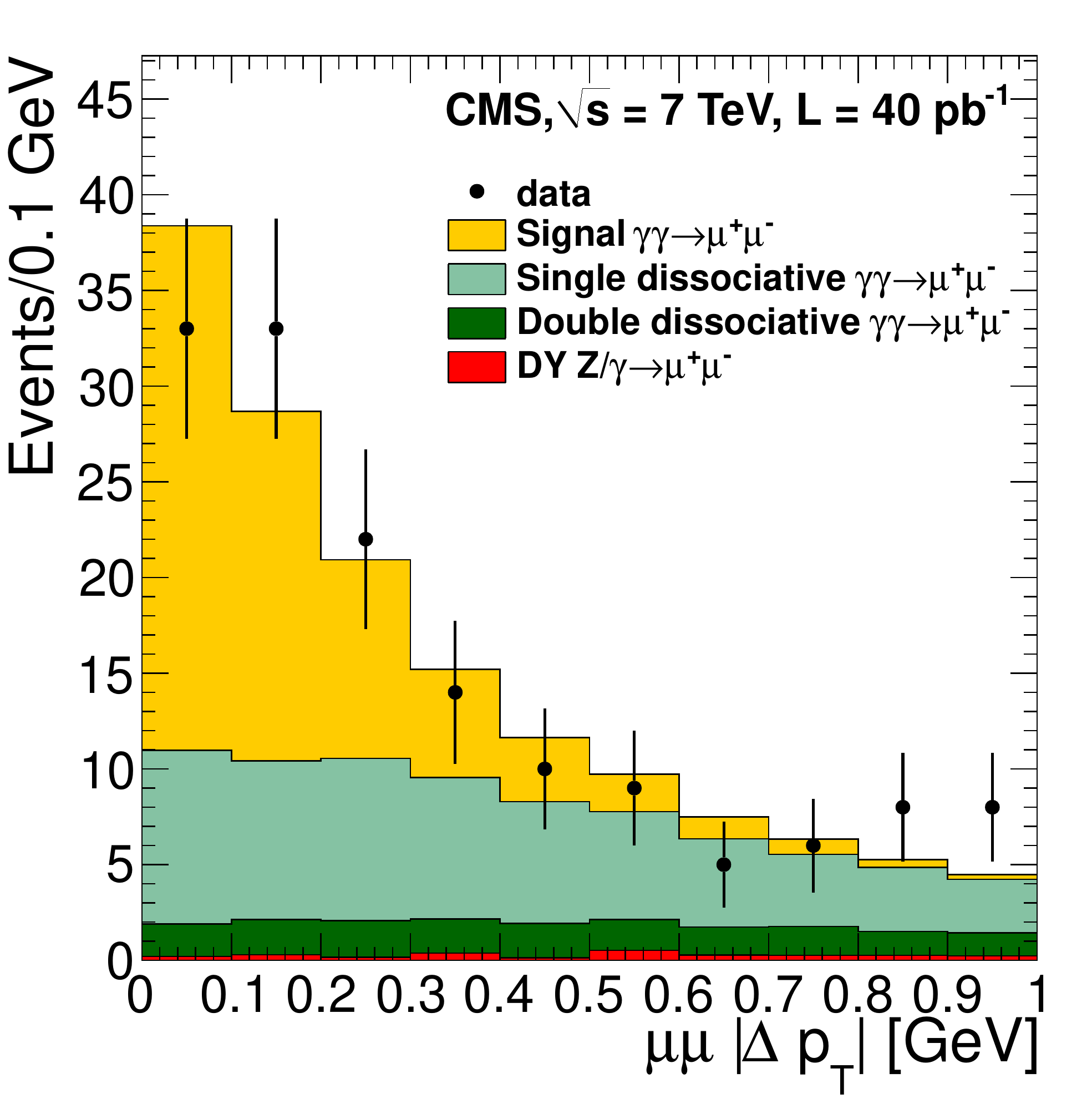}
\includegraphics[width=0.4\textwidth]{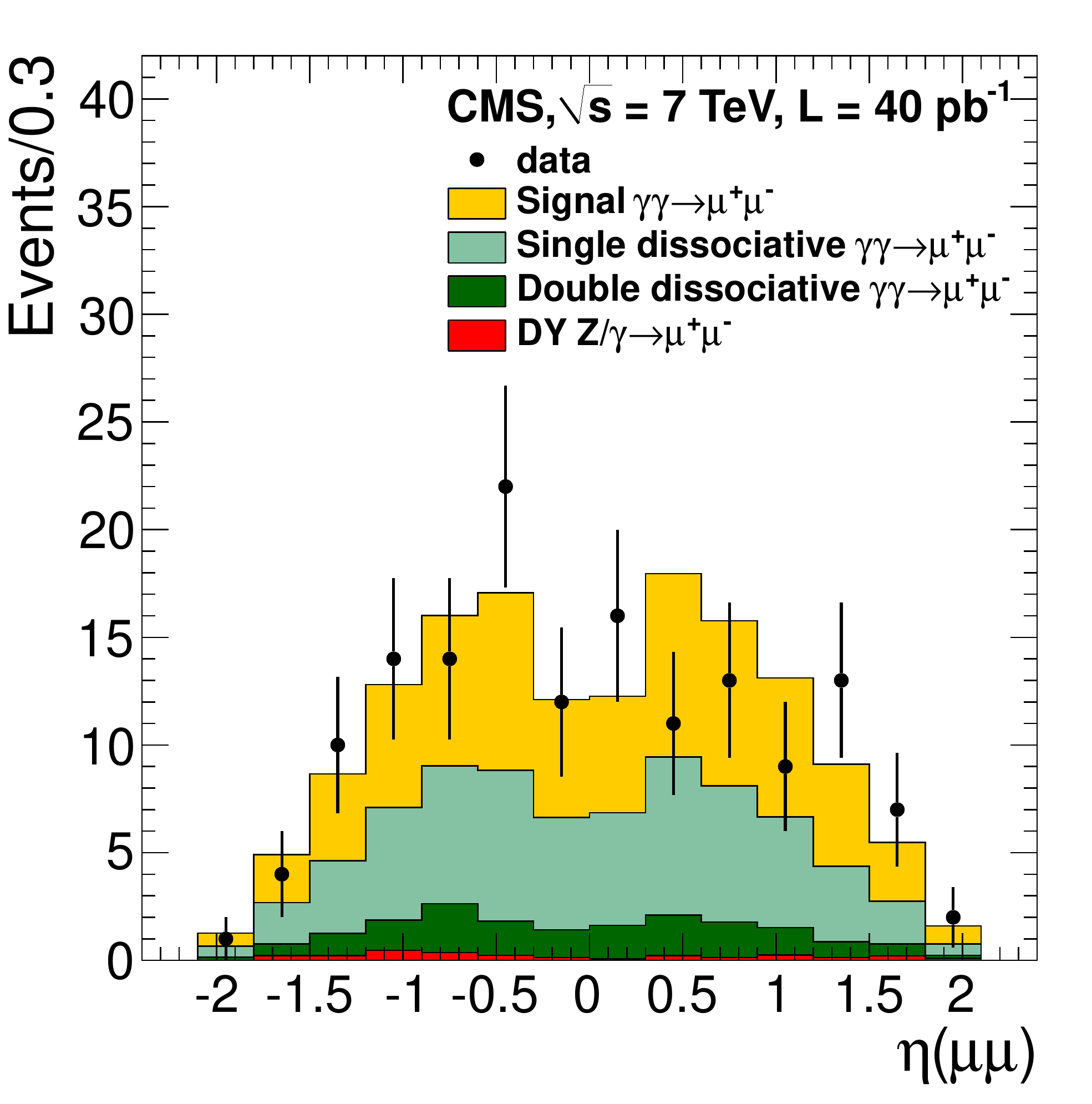}
\caption{Muon pair transverse momentum difference (left) and pair pseudorapidity (right), with all selection criteria applied and
the simulation normalized to the best-fit value. Data are shown as points with statistical error bars, while the histograms
represent the simulated signal (yellow), single (light green) and double (dark green) proton dissociative backgrounds, and DY (red). }
\label{fig:dpt_dphi_n-1}
\end{figure}

\begin{figure}[h!]
\centering
\includegraphics[width=0.4\textwidth]{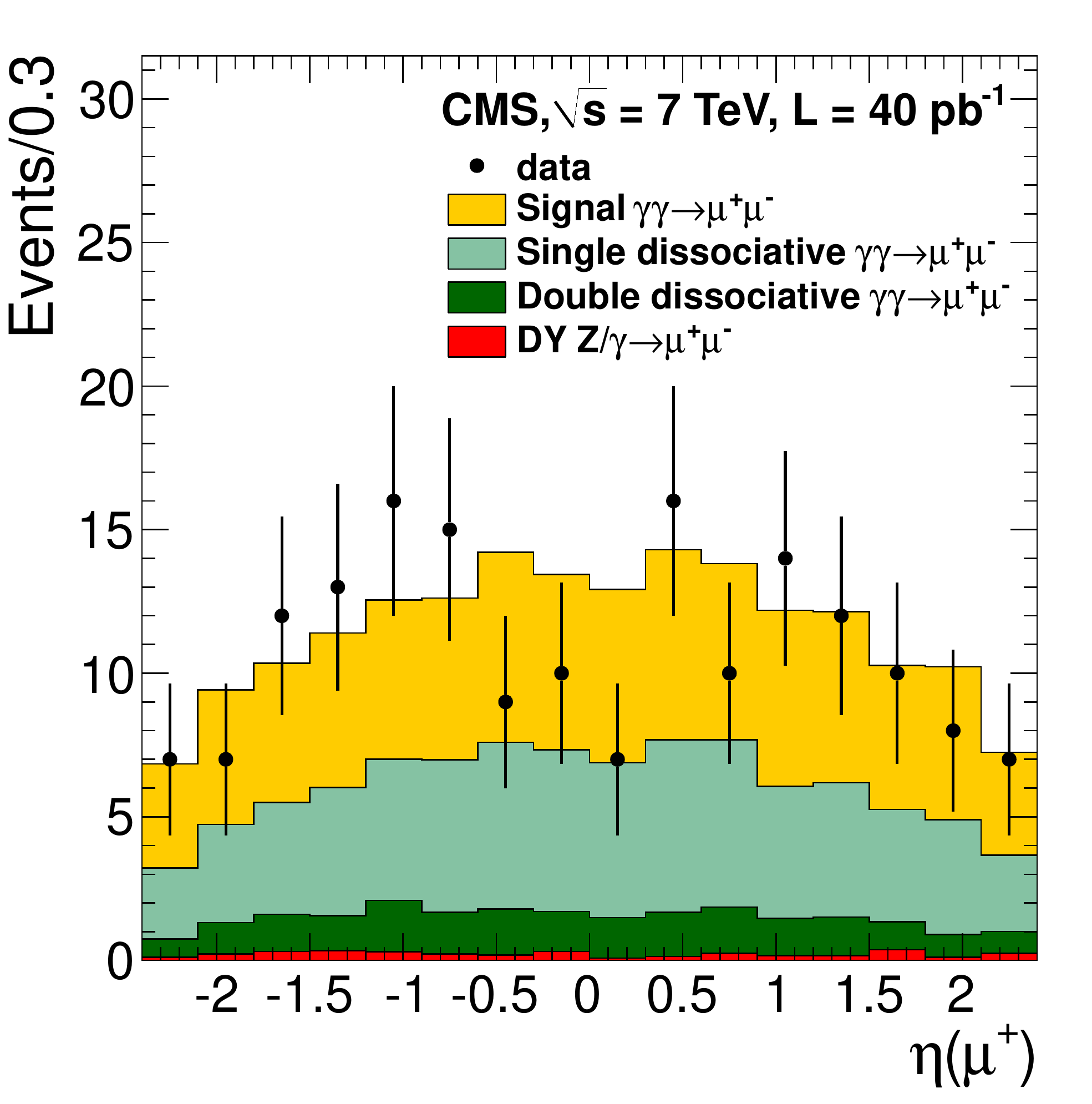}
\includegraphics[width=0.4\textwidth]{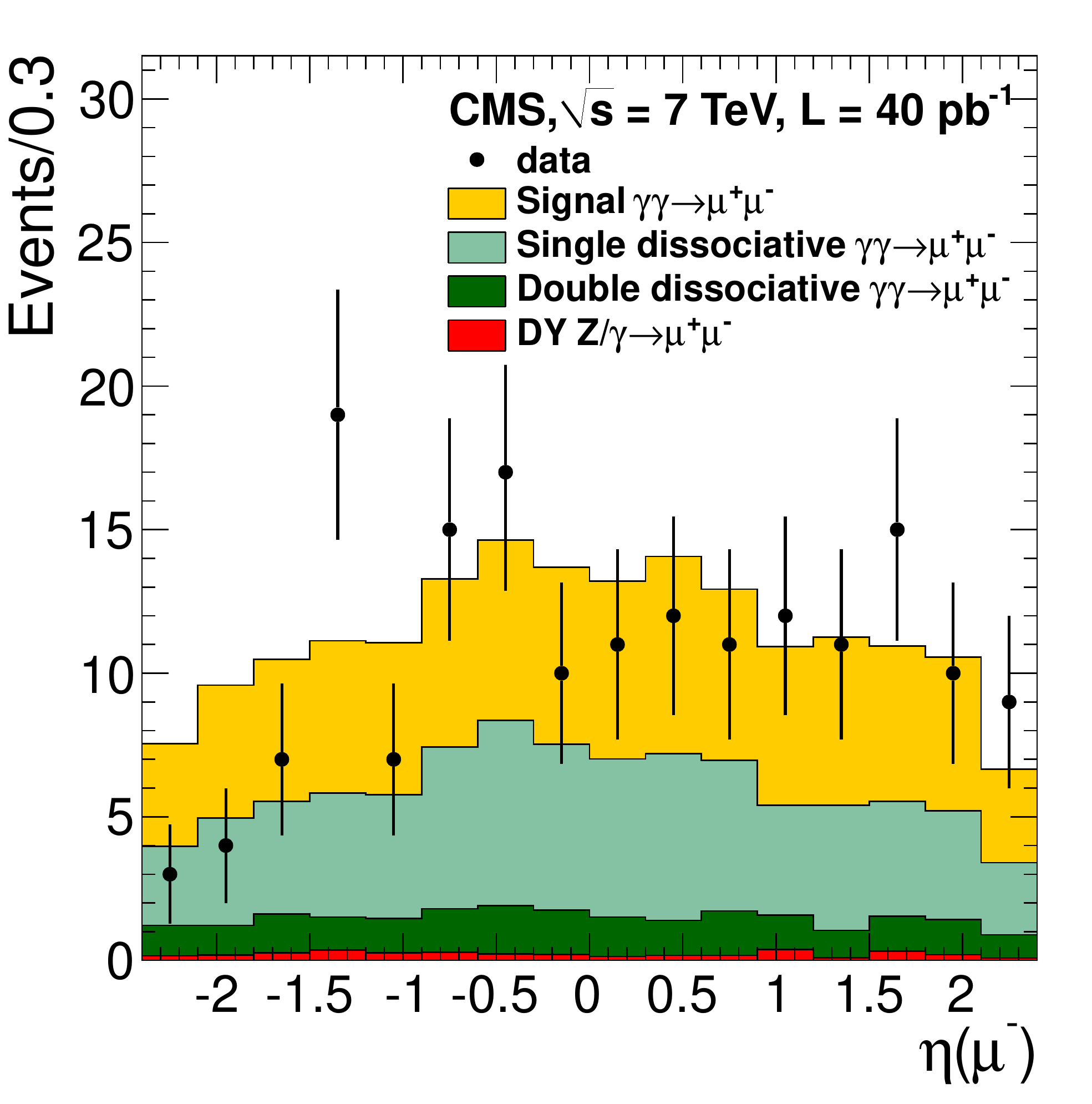}
\caption{Single-muon pseudorapidity distribution with all other selections applied for $\mu^{+}$ (left) and $\mu^{-}$ (right) and the simulation
normalized to the best-fit value. Data are shown as points with statistical error bars, while the histograms
represent the simulated signal (yellow), single (light green) and double (dark green) proton dissociative backgrounds, and DY (red). }
\label{fig:etanm1_all}
\end{figure}

\begin{figure}[h!]
\centering
\includegraphics[width=0.4\textwidth]{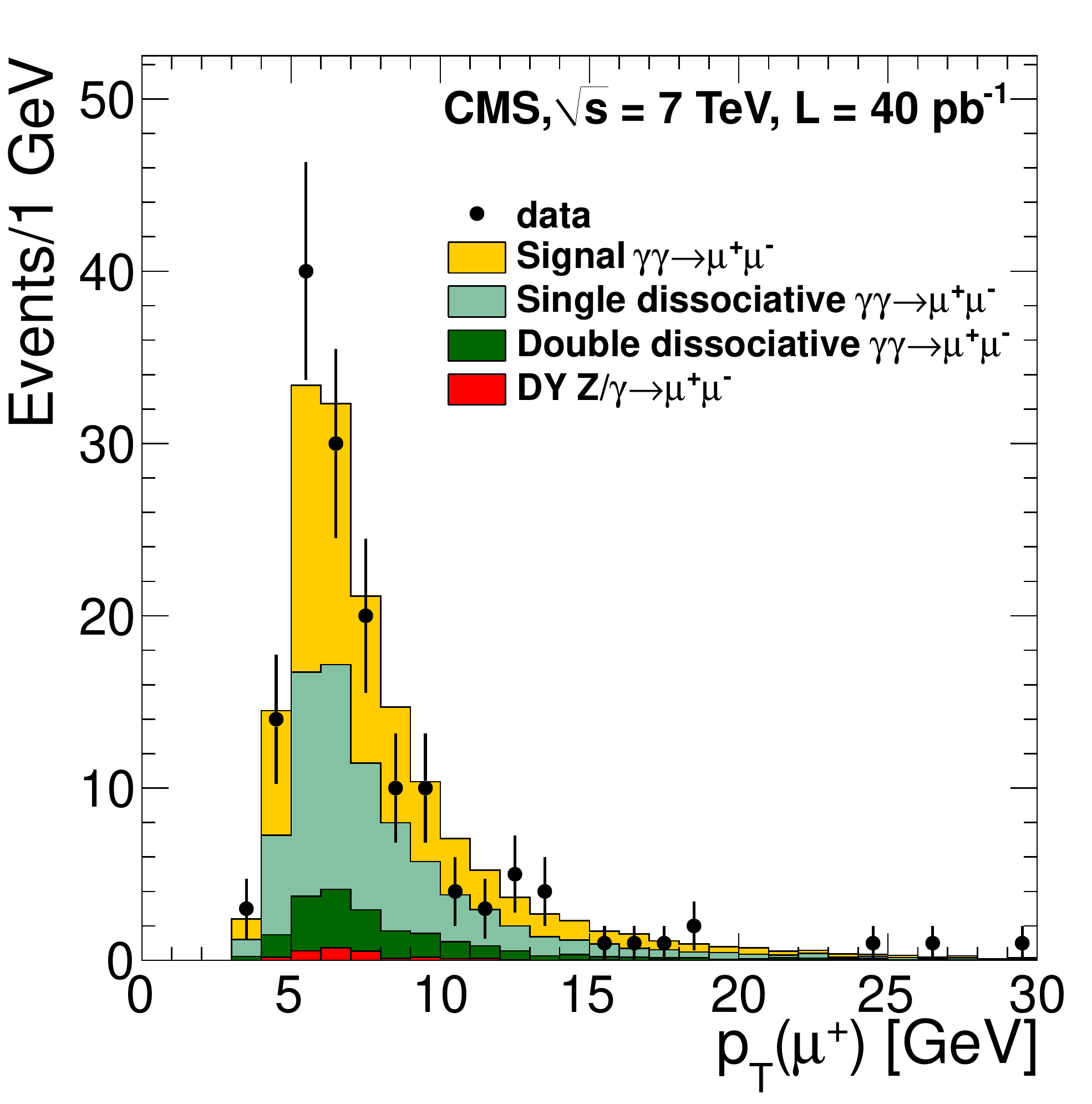}
\includegraphics[width=0.4\textwidth]{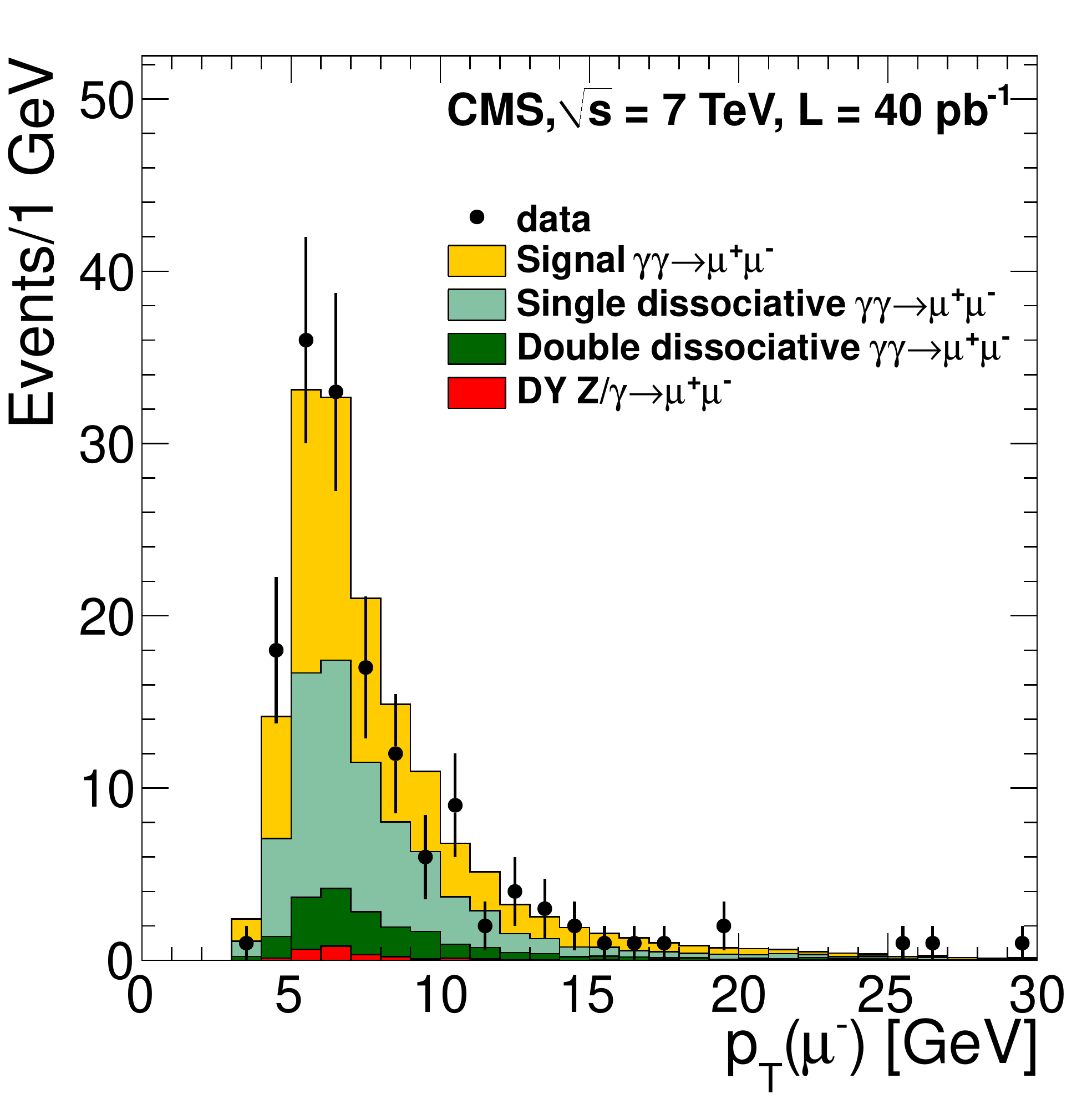}
\caption{Single-muon transverse momentum with all other selections applied for $\mu^{+}$ (left) and $\mu^{-}$ (right) and the
simulation normalized to the best-fit value. Data are shown as points with statistical error bars, while the histograms
represent the simulated signal (yellow), single (light green) and double (dark green) proton dissociative backgrounds, and DY (red). }
\label{fig:ptnm1_all}
\end{figure}

In Fig.~\ref{fig:dpt_dphi_n-1}, the $|\Delta \pt(\Pgmp\Pgmm)|$ and $\eta$ distributions are plotted. In
Figs.~\ref{fig:etanm1_all}--\ref{fig:ptnm1_all}, the data and simulation are similarly compared for the $\pt$ and $\eta$
of single muons passing all other selection requirements. Agreement between the data and simulation is observed in the distributions
of all dimuon and single-muon quantities.

\section{Systematic uncertainties and cross-checks}

Systematic uncertainties related to the pileup efficiency correction, muon trigger and reconstruction
efficiency corrections, momentum scale, LHC crossing angle, and description of the backgrounds in
the fit are considered. The systematic uncertainties related to the muon identification, trigger,
and tracking efficiencies are determined from the statistical uncertainties of the \PJGy and Z
control samples used to derive the corrections. The remaining systematic uncertainties are evaluated by
varying each contribution as described in the following sections, and repeating the fit with the same
three free parameters $R_{El-El}$, $R_{diss-El}$, and the shape correction $a$. The relative difference of the
data-theory signal ratio between the modified and the nominal fit result is taken as a systematic uncertainty.

\subsection{Pileup correction systematic uncertainties}

Charged tracks from pileup interactions more than 2.0\unit{mm} from the dimuon vertex may induce
a signal inefficiency, if they are misreconstructed to originate from within the 2.0\unit{mm}
veto window. The $\eta$-dependent single-track impact parameter resolution in CMS has been measured to be less than 0.2\unit{mm} in the
transverse direction, and less than 1.0\unit{mm} in the longitudinal direction~\cite{bib-vertex}.
The track-veto efficiency is studied in zero-bias data by varying the nominal 2.0\unit{mm} veto distance
from 1.0 to 3.0\unit{mm}. The maximum relative variation is found to be 3.6\%, when enlarging the veto size
to 3.0\unit{mm}. In addition, the veto is modified to use high-quality tracks having at least seven consecutive
layers hit in the tracker, in place of the default veto based on all charged tracks. This is found to result in a
2.5\% variation in the signal yield, which is taken as a systematic uncertainty.

As a further check, the same variations are applied to the selected sample of dimuon events,
removing the $\PgU$ mass cut $m<11.5\GeV$ to increase the statistics with photo-produced
exclusive upsilon events. The change in the number of events selected in the dimuon sample is found to be
consistent with the expectation from the zero-bias sample.

\subsection{Muon efficiencies and momentum scale}

The statistical uncertainty on the muon efficiency correction is evaluated by performing a fast Monte Carlo study in
which each single-muon correction evaluated from the tag-and-probe study is varied independently using a Gaussian distribution having a width
equal to the measured uncertainty. The r.m.s. of the distribution of the resulting variations in the overall dimuon efficiency correction is taken as the
systematic uncertainty. From 1000 pseudo-experiments, this results in an uncertainty of 0.8\%.
In addition, we study the effect of correlations in the dimuon efficiency. The tag-and-probe study is only sensitive to single-muon efficiencies. Since we
take the dimuon efficiency as the product of the single-muon efficiencies, the effect of correlations in the efficiency are not modeled. To
evaluate the size of this effect, the efficiency corrections are computed after removing events in the \PJGy control sample in which the two muons
bend towards each other in the $r$-$\phi$ plane, potentially becoming very close or overlapping. Such events may introduce larger correlations
in the efficiency of the dimuon pair than would be present in the well separated signal muons. Repeating the signal extraction with
this change results in a relative difference of 0.7\% from the nominal efficiency, which is taken as a systematic uncertainty.

Using studies of the muon momentum scale derived from $\cPZ \rightarrow \Pgmp\Pgmm$~\cite{CMSWZ}, the muon $\pt$ is shifted
by the observed $\pt$-dependent bias, and the nominal fit is performed again. The resulting relative change in the signal
yield is 0.1\%, which is taken as a systematic uncertainty. As a cross-check using a sample kinematically closer to the signal, we apply
all the selections except for the veto on the $\PgU$ mass region, and perform a fit to the $\PgUa$ resonance. The resulting mass is consistent
with the PDG value~\cite{PDG}, within an uncertainty of 20\MeV. Applying a 20\MeV shift to the mass and $\pt$ scales of the data and
performing the fit again results in no change from the nominal efficiency.

\subsection{Vertexing and tracking efficiencies}

Since the study described in Section~4.4 shows no significant difference in the vertexing efficiency
between data and simulation, the 0.1\% statistical uncertainty of the measurement in data is
taken as a systematic uncertainty. For the tracking efficiency, the difference between data and
simulation is applied as a single correction without binning in $\pt$ or $\eta$. The
statistical uncertainty of 0.1\% on the correction for the dimuon is taken as a systematic uncertainty.

\subsection{Crossing angle}

The non-zero crossing angle of the LHC beams leads to a boost of the dimuon system in the $x$ direction.
Consequently, the $\pt$ of the pair is over-estimated by a few \MeV, especially for high-mass
dimuon events. This effect is estimated by applying a correction for the Lorentz boost, using a
half-angle of $100\,\mu$rad in the $x$-$z$ plane. This results in a 1.0\% variation from the nominal fit
value, and is taken as an additional systematic uncertainty.

\subsection{Fit stability}

Checks of the fit stability are performed by testing different bin widths and fit ranges. Starting from the nominal
number of 20 bins in the range 0--3\GeV, variations in the bin width from 0.1 to 0.2\GeV and fit range $[0, 2]$ to $[0, 4]$\GeV show
deviations by at most 3.3\% with respect to the nominal yield. The fit bias is studied by performing a series of Monte Carlo
pseudo-experiments for different input values of the signal and proton-dissociation yields, using events drawn from the fully simulated
samples. The means of the pull distributions are found to be consistent with zero. Since the pseudo-experiments with the nominal
binning and fit range show no significant bias, no additional systematics are assigned in this case.

\subsection{Backgrounds}

The yields of the double-proton dissociation and DY contributions are fixed in the nominal fit. To estimate the systematic uncertainty from this
constraint, the fit is repeated with each of these varied independently by a factor of 2. The resulting changes in the fitted signal yield are
0.9\% and 0.4\%, respectively, where because of the similar shapes of the single and double proton dissociation components, this variation is
partly absorbed into the fitted single-proton dissociation yield. As a cross-check of this procedure, the $|\Delta \pt(\Pgmp\Pgmm)|$
and $1 -|\Delta \phi(\Pgmp\Pgmm)/\pi|$ requirements are inverted to select samples of events expected to be dominated by double-proton dissociation
and DY backgrounds. The agreement between data and simulation in these regions is found to be within the factor of 2 used as a systematic variation.

The possibility of a large contamination from cosmic-ray muons, which may fake a signal since they will not
be correlated with other tracks in the event, is studied by comparing the vertex position and three-dimensional opening
angle in data and simulations of collision backgrounds. A total of three events fail the vertex position selection in data,
after all other selection criteria are applied. All three also fail the opening angle selection, which is consistent with the
expected signature from cosmic muons. We conclude that the opening angle requirement effectively rejects cosmic muons, and do not assign
a systematic uncertainty for this possible contamination.

A similar check for contamination from beam-halo muons is performed by applying the nominal analysis selection to non-collision
events triggered by the presence of a single beam. Within the limited statistics, zero events pass all the analysis selections, and
therefore no additional systematic uncertainty is assigned in this case.

\subsection{Summary of systematic uncertainties}

The individual variations in the definition of the track-veto are taken as correlated uncertainties, with the largest variation
taken as a contribution to the systematic uncertainty. The largest variation related to the track quality, obtained when requiring high-purity
tracks with $>10$ hits instead of the nominal value of $>3$ hits, is also taken as a contribution. The larger variation
resulting from increasing or decreasing the double-proton dissociation background normalization by a factor of 2, and
the larger variation resulting from increasing or decreasing the DY background normalization by a factor of 2, are
each taken as contributions to the systematic uncertainty. The variation in the crossing angle, muon identification and trigger efficiencies, tracking
efficiency, bias due to correlations in the \PJGy control sample, and vertexing efficiency are treated as
uncorrelated uncertainties. Summing quadratically all uncorrelated contributions gives an overall relative systematic
uncertainty of 4.8\% on the signal yield (Table~\ref{tab:trkvetosystfit}).

\begin{table}[h!]
\begin{center}
\caption{Relative systematic uncertainties.}
\begin{tabular}{l|c}
\hline
Selection & Variation from nominal yield \\
\hline
Track veto criteria                                      & 3.6\%  \\
Track quality                                        & 2.5\%   \\
DY background                                        & 0.4\%  \\
Double-proton dissociation background                     & 0.9\%  \\
Crossing angle                                       & 1.0\% \\
Tracking efficiency                                  & 0.1\% \\
Vertexing efficiency                                 & 0.1\% \\
Momentum scale                                       & 0.1\% \\
Efficiency correlations in \PJGy control sample   & 0.7\% \\
Muon and trigger efficiency statistical uncertainty  & 0.8\% \\
\hline
Total                                                & 4.8\% \\
\hline
\end{tabular}
\label{tab:trkvetosystfit}
\end{center}
\end{table}

\section{Results}

For muon pairs with invariant mass greater than 11.5\GeV, single-muon transverse momentum $\pt(\mu) > 4\GeV$, and
single-muon pseudorapidity in the range $|\eta(\mu)| < 2.1$, 148 events pass all selections. Approximately half of these are
ascribed to fully exclusive (elastic) production. The number of events expected from Monte Carlo simulation of signal,
proton dissociation, and DY backgrounds for an integrated luminosity of 40\pbinv is 184.

The resulting visible cross section from a fit to the $\pt(\Pgmp\Pgmm)$ distribution is
$\sigma(\Pp\Pp \rightarrow \Pp \Pgmp\Pgmm \Pp) = 3.38^{+0.58}_{-0.55}~(\text{stat.}) \pm 0.16~(\text{syst.}) \pm 0.14~(\text{lumi.})\unit{pb}$, and the corresponding data-theory signal ratio is $0.83^{+0.14}_{-0.13}~(\text{stat.}) \pm 0.04~(\text{syst.}) \pm 0.03~(\text{lumi.})$,
where the statistical uncertainties are strongly correlated with the single-proton dissociation background.

\section{Summary}

A measurement is reported of the exclusive two-photon production of muon pairs, $\Pp\Pp \rightarrow \Pp \mu^+\mu^- \Pp$,
in a 40\pbinv sample of proton-proton collisions collected at $\sqrt{s}= 7$\TeV during 2010 at the LHC. The measured cross section

\begin{eqnarray}
  \sigma(\Pp\Pp \rightarrow \Pp \Pgmp\Pgmm \Pp) = 3.38^{+0.58}_{-0.55}~(\text{stat.}) \pm 0.16~(\text{syst.}) \pm 0.14~(\text{lumi.})~\unit{pb} \nonumber,
\end{eqnarray}

is consistent with the predicted value, and the characteristic distributions of the muon pairs produced via $\Pgg\Pgg$ fusion, such as
the pair acoplanarity and transverse momentum, are well described by the full simulation using the matrix-element event generator \textsc{Lpair}.
The detection efficiencies are determined from control samples in data, including corrections for the significant event pileup.
The signal yield is correlated with the dominant background from two-photon production with proton dissociation, for which the
current estimate from a fit to the $\pt(\Pgmp\Pgmm)$ distribution can be improved with additional data. The
efficiency for the exclusivity selection is above 90\% in the full data sample collected by CMS during the 2010 LHC run. With
increasing instantaneous luminosity this efficiency will decrease, but without possible improvements to the selection remains
above 60\% with up to 8 additional pileup vertices. Since the process may be calculated reliably in the framework of QED, within
uncertainties associated with the proton form factor, this represents a first step towards a complementary luminosity
measurement, and a reference for other exclusive production measurements to be performed with pileup.

\section*{Acknowledgments}
\hyphenation{Bundes-ministerium Forschungs-gemeinschaft Forschungs-zentren} We wish to congratulate our colleagues in the CERN accelerator departments for the excellent performance of the LHC machine. We thank the technical and administrative staff at CERN and other CMS institutes. This work was supported by the Austrian Federal Ministry of Science and Research; the Belgium Fonds de la Recherche Scientifique, and Fonds voor Wetenschappelijk Onderzoek; the Brazilian Funding Agencies (CNPq, CAPES, FAPERJ, and FAPESP); the Bulgarian Ministry of Education and Science; CERN; the Chinese Academy of Sciences, Ministry of Science and Technology, and National Natural Science Foundation of China; the Colombian Funding Agency (COLCIENCIAS); the Croatian Ministry of Science, Education and Sport; the Research Promotion Foundation, Cyprus; the Estonian Academy of Sciences and NICPB; the Academy of Finland, Finnish Ministry of Education and Culture, and Helsinki Institute of Physics; the Institut National de Physique Nucl\'eaire et de Physique des Particules~/~CNRS, and Commissariat \`a l'\'Energie Atomique et aux \'Energies Alternatives~/~CEA, France; the Bundesministerium f\"ur Bildung und Forschung, Deutsche Forschungsgemeinschaft, and Helmholtz-Gemeinschaft Deutscher Forschungszentren, Germany; the General Secretariat for Research and Technology, Greece; the National Scientific Research Foundation, and National Office for Research and Technology, Hungary; the Department of Atomic Energy and the Department of Science and Technology, India; the Institute for Studies in Theoretical Physics and Mathematics, Iran; the Science Foundation, Ireland; the Istituto Nazionale di Fisica Nucleare, Italy; the Korean Ministry of Education, Science and Technology and the World Class University program of NRF, Korea; the Lithuanian Academy of Sciences; the Mexican Funding Agencies (CINVESTAV, CONACYT, SEP, and UASLP-FAI); the Ministry of Science and Innovation, New Zealand; the Pakistan Atomic Energy Commission; the State Commission for Scientific Research, Poland; the Funda\c{c}\~ao para a Ci\^encia e a Tecnologia, Portugal; JINR (Armenia, Belarus, Georgia, Ukraine, Uzbekistan); the Ministry of Science and Technologies of the Russian Federation, the Russian Ministry of Atomic Energy and the Russian Foundation for Basic Research; the Ministry of Science and Technological Development of Serbia; the Ministerio de Ciencia e Innovaci\'on, and Programa Consolider-Ingenio 2010, Spain; the Swiss Funding Agencies (ETH Board, ETH Zurich, PSI, SNF, UniZH, Canton Zurich, and SER); the National Science Council, Taipei; the Scientific and Technical Research Council of Turkey, and Turkish Atomic Energy Authority; the Science and Technology Facilities Council, UK; the US Department of Energy, and the US National Science Foundation.

Individuals have received support from the Marie-Curie programme and the European Research Council (European Union); the Leventis Foundation; the A. P. Sloan Foundation; the Alexander von Humboldt Foundation; the Belgian Federal Science Policy Office; the Fonds pour la Formation \`a la Recherche dans l'Industrie et dans l'Agriculture (FRIA-Belgium); the Agentschap voor Innovatie door Wetenschap en Technologie (IWT-Belgium); and the Council of Science and Industrial Research, India.

\bibliography{auto_generated}   

\providecommand{\href}[2]{#2}\begingroup\raggedright\begin{thebibliography}{10}%
\makeatletter
\providecommand{\hrefCMSnoop }[0]{\@secondoftwo}%
\makeatother

\bibitem{Budnev1}
\hrefCMSnoop {} {V.~M. Budnev, I.~F. Ginzburg, G.~V. Meledin{ et~al.}, ``{The
  process p p $\rightarrow$ p p e+ e- and the possibility of its calculation by
  means of quantum electrodynamics only}'',} \textit{ Nucl. Phys B} \textbf{
  63} (1973) 519.
\href{http://dx.doi.org/10.1016/0550-3213(73)90162-4}{\texttt{
  doi:10.1016/0550-3213(73)90162-4}}.

\bibitem{lumi2}
\hrefCMSnoop {} {A.~G. Shamov and V.~I. Telnov, ``{Precision luminosity
  measurement at LHC using two-photon production of $\mu^{+} \mu^{-}$
  pairs}'',} \textit{ Nucl. Instrum. Meth. A} \textbf{ 494} (2002) 51,
  \href{http://www.arXiv.org/abs/hep-ex/0207095}{\texttt{
  arXiv:hep-ex/0207095}}.
\href{http://dx.doi.org/10.1016/S0168-9002(02)01444-4}{\texttt{
  doi:10.1016/S0168-9002(02)01444-4}}.

\bibitem{lumi1}
\hrefCMSnoop {} {V.~A. Khoze, A.~D. Martin, R.~Orava{ et~al.}, ``Luminosity
  monitors at the {LHC}'',} \textit{ Eur. Phys. J.} \textbf{ C19} (2001) 313,
  \href{http://www.arXiv.org/abs/hep-ph/0010163}{\texttt{
  arXiv:hep-ph/0010163}}.
\href{http://dx.doi.org/10.1007/s100520100616}{\texttt{
  doi:10.1007/s100520100616}}.

\bibitem{Abulencia}
\hrefCMSnoop {} {{ CDF} Collaboration, ``Observation of exclusive
  electron-positron production in hadron-hadron collisions'',} \textit{ Phys.
  Rev. Lett.} \textbf{ 98} (2007) 112001,
  \href{http://www.arXiv.org/abs/hep-ex/0611040}{\texttt{
  arXiv:hep-ex/0611040}}.
\href{http://dx.doi.org/10.1103/PhysRevLett.98.112001}{\texttt{
  doi:10.1103/PhysRevLett.98.112001}}.

\bibitem{CDFZ}
\hrefCMSnoop {} {{ CDF} Collaboration, ``Search for exclusive {$Z$}-boson
  production and observation of high mass $p\bar{p} \to \gamma \gamma \to
  p\ell^{+}\ell^{-}\bar{p}$ events in $p\bar{p}$ collisions at $\sqrt{s}$ =
  1.96 {TeV}'',} \textit{ Phys. Rev. Lett.} \textbf{ 102} (2009) 222002,
  \href{http://www.arXiv.org/abs/0902.2816}{\texttt{ arXiv:0902.2816}}.
\href{http://dx.doi.org/10.1103/PhysRevLett.102.222002}{\texttt{
  doi:10.1103/PhysRevLett.102.222002}}.

\bibitem{CDFchic}
\hrefCMSnoop {} {{ CDF} Collaboration, ``Observation of exclusive charmonium
  production and $\gamma\gamma \to \mu^+\mu^-$ in $p\bar{p}$ collisions at
  $\sqrt{s} = 1.96$ {TeV}'',} \textit{ Phys. Rev. Lett.} \textbf{ 102} (2009)
  242001, \href{http://www.arXiv.org/abs/0902.1271}{\texttt{ arXiv:0902.1271}}.
\href{http://dx.doi.org/10.1103/PhysRevLett.102.242001}{\texttt{
  doi:10.1103/PhysRevLett.102.242001}}.

\bibitem{LUMIDP}
\href {http://cdsweb.cern.ch/record/1335668} {{ CMS} Collaboration, ``Absolute
  luminosity normalization'',} CMS Detector Performance Study CMS-DP-2010-002,
  (2010).

\bibitem{JINST}
\hrefCMSnoop {} {S.~Chatrchyan {et~al.}, ``The {CMS} experiment at the {CERN}
  {LHC}'',} \textit{ JINST} \textbf{ 03} (2008) S08004.
\href{http://dx.doi.org/10.1088/1748-0221/3/08/S08004}{\texttt{
  doi:10.1088/1748-0221/3/08/S08004}}.

\bibitem{Vermaseren}
\hrefCMSnoop {} {J.~A.~M. Vermaseren, ``{Two Photon Processes at Very High
  Energies}'',} \textit{ Nucl. Phys. B} \textbf{ 229} (1983) 347.
\href{http://dx.doi.org/10.1016/0550-3213(83)90336-X}{\texttt{
  doi:10.1016/0550-3213(83)90336-X}}.

\bibitem{Baranov}
\hrefCMSnoop {} {S.~P. Baranov {et~al.}, ``{LPAIR} - A generator for lepton
  pair production'',} in \textit{ Proceedings of Physics at HERA}, p.~1478.
\newblock October, 1991.

\bibitem{lund}
\hrefCMSnoop {} {B.~Andersson, G.~Gustafson, G.~Ingelman{ et~al.}, ``Parton
  fragmentation and string dynamics'',} \textit{ Phys. Rep.} \textbf{ 97}
  (1983) 31.
\href{http://dx.doi.org/10.1016/0370-1573(83)90080-7}{\texttt{
  doi:10.1016/0370-1573(83)90080-7}}.

\bibitem{jetset}
\hrefCMSnoop {} {T.~Sj{\"o}strand, ``{High-energy physics event generation with
  PYTHIA 5.7 and JETSET 7.4}'',} \textit{ Comput. Phys. Commun.} \textbf{ 82}
  (1994) 74.
\href{http://dx.doi.org/10.1016/0010-4655(94)90132-5}{\texttt{
  doi:10.1016/0010-4655(94)90132-5}}.

\bibitem{brasse}
\hrefCMSnoop {} {F.~W. Brasse, W.~Flauger, J.~Gayler{ et~al.},
  ``Parametrization of the $q^2$ dependence of $\gamma_V p$ total cross
  sections in the resonance region'',} \textit{ Nucl. Phys. B} \textbf{ 110}
  (1976) 413. \href{http://dx.doi.org/10.1016/0550-3213(76)90231-5}{\texttt{
  doi:10.1016/0550-3213(76)90231-5}}.

\bibitem{suri}
\hrefCMSnoop {} {A.~Suri and D.~R. Yennie, ``{The space-time phenomenology of
  photon absorption and inelastic electron scattering}'',} \textit{ Annals
  Physics} \textbf{ 72} (1972) 43.
  \href{http://dx.doi.org/10.1016/0003-4916(72)90242-4}{\texttt{
  doi:10.1016/0003-4916(72)90242-4}}.

\bibitem{pythia6}
\hrefCMSnoop {} {T.~Sj{\"o}strand, L.~L{\"o}nnblad, and S.~Mrenna, ``{PYTHIA}
  6.2: Physics and manual'',} (2001).
\href{http://www.arXiv.org/abs/hep-ph/0108264}{\texttt{ arXiv:hep-ph/0108264}}.

\bibitem{tuneZ2}
\hrefCMSnoop {} {R.~Field, ``Early LHC Underlying Event Data - Finding and
  Surprises'',} (2010). \href{http://www.arXiv.org/abs/1010.3558}{\texttt{
  arXiv:1010.3558}}.

\bibitem{geant}
\hrefCMSnoop {} {S.~Agostinelli {et~al.}, ``GEANT4: A Simulation toolkit'',}
  \textit{ Nucl. Instrum. Meth. A} \textbf{ 506} (2003) 250.
\href{http://dx.doi.org/10.1016/S0168-9002(03)01368-8}{\texttt{
  doi:10.1016/S0168-9002(03)01368-8}}.

\bibitem{CRAFTMUONS}
\hrefCMSnoop {} {{ CMS} Collaboration, ``{Performance of CMS Muon
  Reconstruction in Cosmic-Ray Events}'',} \textit{ JINST} \textbf{ 5} (2010)
  T03022, \href{http://www.arXiv.org/abs/0911.4994}{\texttt{ arXiv:0911.4994}}.
\href{http://dx.doi.org/10.1088/1748-0221/5/03/T03022}{\texttt{
  doi:10.1088/1748-0221/5/03/T03022}}.

\bibitem{bib-trackingefficiency}
\href {http://cdsweb.cern.ch/record/1279139} {{ CMS} Collaboration,
  ``Measurement of Tracking Efficiency'',} CMS Physics Analysis Summary
  CMS-PAS-TRK-10-002, (2010).

\bibitem{bib-vertex}
\href {http://cdsweb.cern.ch/record/1279383} {{ CMS} Collaboration, ``Tracking
  and Primary Vertex Results in First 7 {TeV} Collisions'',} CMS Physics
  Analysis Summary CMS-PAS-TRK-10-005, (2010).

\bibitem{bib-adaptivevtx}
\href {http://cdsweb.cern.ch/record/1027031} {{ CMS} Collaboration, ``Adaptive
  Vertex Fitting'',} CMS Note CMS-NOTE-07-008, (2007).

\bibitem{MUPOGICHEPPAS}
\href {http://cdsweb.cern.ch/record/1279140} {{ CMS} Collaboration,
  ``Performance of muon identification in pp collisions at $\sqrt{s}$ = 7
  {TeV}'',} CMS Physics Analysis Summary CMS-PAS-MUO-10-002, (2010).

\bibitem{CMSWZ}
\hrefCMSnoop {} {{ CMS} Collaboration, ``{Measurement of the Inclusive W and Z
  Production Cross Sections in p p Collisions at sqrt(s) = 7 TeV with the CMS
  experiment}'',} \textit{ JHEP} \textbf{ 10} (2011) 132,
  \href{http://www.arXiv.org/abs/1107.4789}{\texttt{ arXiv:1107.4789}}.
\href{http://dx.doi.org/10.1007/JHEP10(2011)132}{\texttt{
  doi:10.1007/JHEP10(2011)132}}.

\bibitem{Kalman}
\hrefCMSnoop {} {R.~Fruhwirth, ``{Application of Kalman filtering to track and
  vertex fitting}'',} \textit{ Nucl. Instrum. Meth. A} \textbf{ 262} (1987)
  444--450.
\href{http://dx.doi.org/10.1016/0168-9002(87)90887-4}{\texttt{
  doi:10.1016/0168-9002(87)90887-4}}.

\bibitem{minuit}
\hrefCMSnoop {} {F.~James and M.~Roos, ``{Minuit: A System for Function
  Minimization and Analysis of the Parameter Errors and Correlations}'',}
  \textit{ Comput. Phys. Commun.} \textbf{ 10} (1975) 343.
\href{http://dx.doi.org/10.1016/0010-4655(75)90039-9}{\texttt{
  doi:10.1016/0010-4655(75)90039-9}}.

\bibitem{PDG}
\hrefCMSnoop {} {{ Particle Data Group} Collaboration, ``Review of particle
  physics'',} \textit{ J. Phys G} \textbf{ 37} (2010) 075021.
\href{http://dx.doi.org/10.1088/0954-3899/37/7A/075021}{\texttt{
  doi:10.1088/0954-3899/37/7A/075021}}.

\end{thebibliography}\endgroup
\clearpage
\newpage
\appendix
\cleardoublepage \appendix\section{The CMS Collaboration \label{app:collab}}\begin{sloppypar}\hyphenpenalty=5000\widowpenalty=500\clubpenalty=5000\textbf{Yerevan Physics Institute,  Yerevan,  Armenia}\\*[0pt]
S.~Chatrchyan, V.~Khachatryan, A.M.~Sirunyan, A.~Tumasyan
\vskip\cmsinstskip
\textbf{Institut f\"{u}r Hochenergiephysik der OeAW,  Wien,  Austria}\\*[0pt]
W.~Adam, T.~Bergauer, M.~Dragicevic, J.~Er\"{o}, C.~Fabjan, M.~Friedl, R.~Fr\"{u}hwirth, V.M.~Ghete, J.~Hammer\cmsAuthorMark{1}, S.~H\"{a}nsel, M.~Hoch, N.~H\"{o}rmann, J.~Hrubec, M.~Jeitler, W.~Kiesenhofer, M.~Krammer, D.~Liko, I.~Mikulec, M.~Pernicka, B.~Rahbaran, H.~Rohringer, R.~Sch\"{o}fbeck, J.~Strauss, A.~Taurok, F.~Teischinger, C.~Trauner, P.~Wagner, W.~Waltenberger, G.~Walzel, E.~Widl, C.-E.~Wulz
% ...
\vskip\cmsinstskip
% ...
1:~~Also at CERN, European Organization for Nuclear Research, Geneva, Switzerland\\
% ...

\end{sloppypar}
\end{document}